\documentstyle[aps,amsmath,psfig]{revtex}

\def\eq#1{Eq.~(\ref{#1})}
\def\tab#1{Tab.~\ref{#1}}
\def\fig#1{Fig.~\ref{#1}}

\begin{document}

\title{Bending Frustration of Lipid-Water Mesophases Based on 
       Cubic Minimal Surfaces\cite{letter}}
\author{U. S. Schwarz${}^{a,}$\cite{newaddress} and G. Gompper${}^{b}$ \\
${}^{a}$Department of Materials and Interfaces,
Weizmann Institute of Science,
Rehovot 76100, Israel \\
${}^{b}$Institut f\"ur Festk\"orperforschung,
Forschungszentrum J\"ulich,
52425 J\"ulich, Germany}  

\date{\today}

\maketitle

\begin{abstract}
  Inverse bicontinuous cubic phases are ubiquitous in lipid-water
  mixtures and consist of a lipid bilayer forming a cubic minimal
  surface, thereby dividing space into two cubic networks of water
  channels. For small hydrocarbon chain lengths, the monolayers can be 
  modeled as parallel surfaces to a minimal midsurface.  The bending
  energy of the cubic phases is determined by the
  distribution of Gaussian curvature over the minimal midsurfaces
  which we calculate for seven different structures (G, D, P, I-WP,
  C(P), S and F-RD). We show that the free-energy densities of the
  structures G, D and P are considerably lower than those of the other
  investigated structures due to their narrow distribution of Gaussian
  curvature.  The Bonnet transformation between G, D, and P implies that
  these phases coexist along a triple line, which also includes an 
  excess water phase.  Our model includes thermal membrane undulations. 
  Our qualitative predictions remain unchanged when higher order terms in 
  the curvature energy are included. Calculated phase diagrams agree 
  well with the experimental results for 2:1 lauric acid/dilauroyl 
  phosphatidylcholine and water.
\end{abstract}

\section{Introduction}
\label{sec:Introduction}

When dissolved in aqueous solvent, lipids self-assemble into an
amazing variety of different structures as a function of concentration
and temperature \cite{a:font90,a:luzz93,a:sedd95}. Most prominent
is the lamellar phase, which consists of a stack of lipid bilayers
separated by layers of water. In fact, the lipid bilayer constitutes
the basic building block of biological membranes; in eucaryotic cells,
it not only envelopes the cell itself, but also separates its nucleus
and its organelles from the cytosol, and builds up the endoplasmic
reticulum, the Golgi apparatus and the uni-lamellar vesicles which
serve as reaction chambers and transport vehicles.  The predominance
of the lamellar phase at ambient temperatures derives from the fact
that in contrast to surfactants, which usually have large headgroups
and form micelles, lipids have rather bulky hydrocarbon chains. A
so-called \emph{bilayer lipid} has a chain region cross section which
is of similar size as the area per head group; thus the spontaneous
curvature is small and the lipid forms a lamellar phase.

If temperature is increased, the size of the head groups' hydration
shell decreases, spontaneous curvature increases and the lipid
monolayers tend to curve towards the water regions. Since the tendency
of lipid monolayers to curve is frustrated in the lamellar phase, it
is often found to transform into a inverse hexagonal phase which
consists of a two-dimensional array of inverse cylindrical
micelles. Spontaneous curvature can also be increased by changing
molecular architecture, by adding lipids with bulkier chains or by
increasing salt concentration (since salt screens electrostatic
repulsion between charged headgroups) \cite{a:port92}. Lipids with
spontaneous curvature are often called \emph{non-bilayer lipids}.  For
large spontaneous curvature, a cubic phase of inverse spherical
micelles is sometimes observed \cite{a:sedd95}.  The inverse phases
have been denoted \emph{type II (water-in-oil) phases} in contrast to
\emph{type I (oil-in-water) phases} which dominate surfactant phase
behavior \cite{a:luzz68}.  Although there are some lipids which also
form type I phases, we focus here on type II phases since these are
the ones formed by the majority of biologically relevant lipids.

One of the intriguing aspects of the polymorphism of lipid-water
mixtures is the existence of yet another structural type, which often
occurs between the lamellar and the inverse hexagonal phase. Like the
lamellar phase, these \emph{inverse bicontinuous cubic phases} (IBCPs)
consist of lipid bilayers. However, now a {\em single} lipid bilayer extends
throughout the whole of space, and divides it into two disconnected but
interpenetrating labyrinths of cubic symmetry, which both are filled
with water. Thus the structure can also be considered to be built from
water channels (that is inverse cylindrical micelles) as is the
inverse hexagonal phase, only that now the channels meet in vertices with a
coordination number related to the topology of the cubic phase.
Although the IBCPs unite the structural elements of the
lamellar and the inverse hexagonal phases, in contrast to them they are
optically isotropic and highly viscous. Like the inverse hexagonal
phase, the IBCPs are stabilized by spontaneous curvature. Therefore
they are formed by non-bilayer lipids and their stability can be
controlled by changing the lipids' spontaneous curvature.

The generic occurrence of IBCPs in lipid-water mixtures and their
property of dividing space into two interwoven aqueous compartments
lends itself readily to speculations on their biological relevance,
e.g.\ on their possible occurrence in the endoplasmic reticulum or the
Golgi apparatus \cite{a:luzz93,a:sedd95}. A recent survey of a vast
number of TEM pictures suggests that IBCPs are indeed ubiquitous in
biological systems \cite{a:land95c,a:hyde97}. The abundance of
non-bilayer lipids in biological systems is a longstanding puzzle, and
several studies have explored the possible relationship between lipid
polymorphism and biological membrane function (for reviews see
Refs.~\cite{a:epan96,a:luzz97,a:krui97}).  In particular, it was shown that
certain cells like \textit{E. Coli} regulate the relative amount of
bilayer and non-bilayer lipid in order to control membrane morphology
and function of membrane proteins. \textit{E. Coli} mutants lacking
the non-bilayer lipid phosphatidylethanolamine were found to be
severely impaired in regard to protein transport across the plasma
membrane \cite{a:riet95}.  Luzzati and coworkers have pointed out that
if any IBCP had a biological function, it should be stable at
physiological temperatures and in excess water
\cite{a:luzz93,a:luzz97}.  IBCPs are also relevant for biotechnological
applications, e.g.\ they have been used recently as artificial matrix
which enables membrane proteins to crystallize in a three-dimensional
array \cite{a:peba97}.

Since IBCPs consist of a lipid bilayer separating two labyrinths of
water, is it reasonable to assume that the two monolayers are arranged
in a way which is \emph{locally} symmetrical about their mid-surface.
In differential geometry, a surface which curves to both sides in the
same way is known as a \emph{minimal surface}, i.e.\ a surface for
which the mean curvature vanishes at each of its points. Therefore the
mid-surfaces of the IBCP's lipid bilayers are often modeled as
\emph{triply periodic minimal surfaces} (TPMS, sometimes also denoted
IPMS for \emph{infinite periodic minimal surfaces}) of cubic symmetry
\cite{a:auss90,a:ptrs96}.  Indeed it has been confirmed by a thorough
analysis of electron density maps derived from X-ray data that the
mid-surfaces of these structures can be very close to minimal surfaces
\cite{a:luzz93}.  Note that the stronger requirement of the
monolayers' \emph{global} symmetry about the mid-surface leads to a
certain subclass of cubic TPMS which is called \emph{balanced} since
then the two water labyrinths have to be congruent to each other.

\begin{figure}
\begin{center}
\leavevmode 
\psfig{file=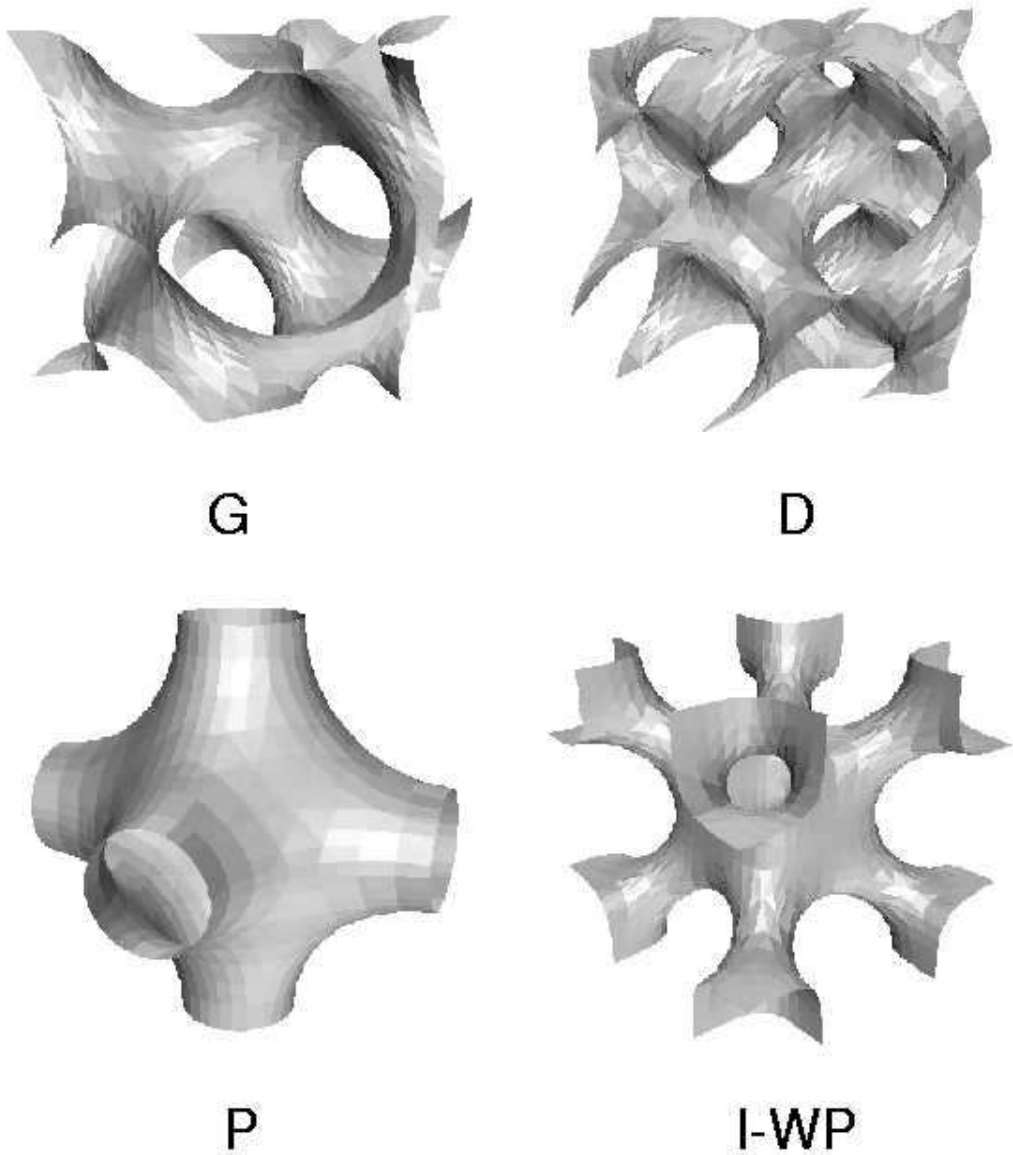,width=0.8\textwidth}
\end{center}
\caption{ One of the two monolayers of P, D, G and I-WP
  for hydrocarbon volume fraction $v=0.5$. These structures have
  space groups $Pm\bar{3}m$, $Fd\bar{3}m$, $I4_132$ and $Im\bar{3}m$,
  respectively.  In the full structures, a second monolayer exists 
  parallel to the one shown. Each monolayer defines a space-percolating network
  filled with water. For the balanced cases P, D and G, the second
  network is congruent to the first; this additional symmetry changes
  the structures' space groups to $Im\bar{3}m$, $Pn\bar{3}m$ and
  $Ia\bar3d$, respectively.  For I-WP, the two networks are different
  and only the monolayer corresponding to the I-network is shown.}
\label{FigureParallelSurfaces1}
\end{figure}

\begin{figure}
\begin{center}
\leavevmode 
\psfig{file=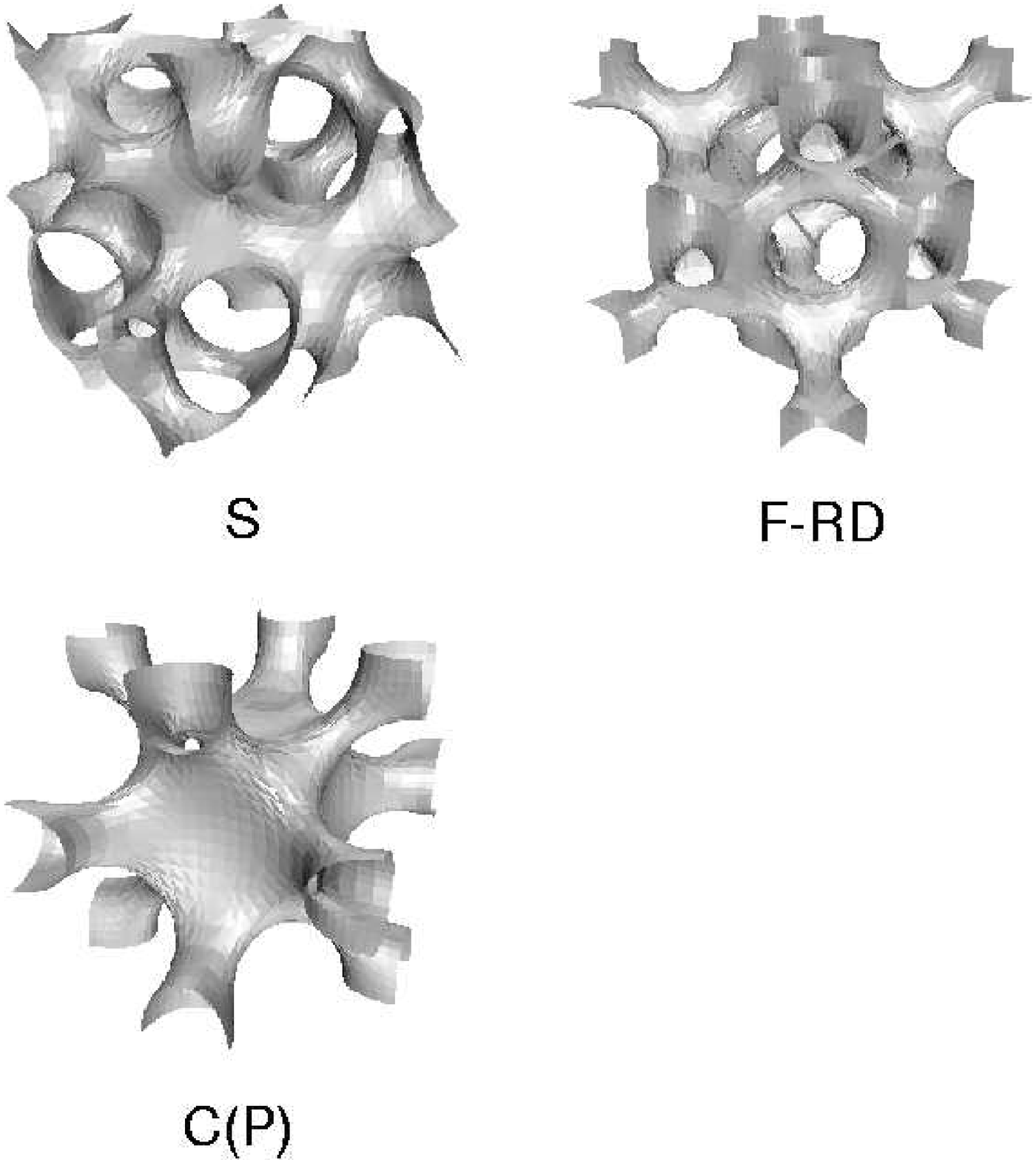,width=0.8\textwidth}
\end{center}
\caption{One of the two monolayers of S, F-RD and C(P)
  for a hydrocarbon volume fraction $v=0.5$. These structures have
  space groups $I\bar43d$, $Fm\bar{3}m$ and $Pm\bar3m$, respectively.
  In the full structures, a second monolayer exists parallel to the
  one shown.  For the balanced cases S and C(P), the second network is
  congruent to the first; this additional symmetry changes the
  structures' space groups to $Im\bar{3}m$ and $Ia\bar3d$,
  respectively.  For F-RD, the two networks are different and only the
  monolayer corresponding to the F-network is shown.}
\label{FigureParallelSurfaces2}
\end{figure}

Since the seminal work of Schoen \cite{a:scho70}, it is known that
there exists a large variety of different cubic TPMS. Before Schoen's
work, only P, D and C(P) were known from the 19th century work of
Schwarz and his students; then Schoen described G, F-RD, I-WP, O,C-TO
and C(D). More cubic TPMS have been found later by Karcher and
Polthier \cite{a:karc96}, Fischer and Koch \cite{a:fisc96} and others,
but most of them seem to be too complicated as to be of physical
relevance. In Figs.~\ref{FigureParallelSurfaces1} and
\ref{FigureParallelSurfaces2} we depict surfaces corresponding to 
the seven different TPMS treated in this work.  Until now, only G, D
and P have been established to be realized as IBCP in lipid-water
mixtures \cite{a:font90,a:luzz93,a:sedd95}. One system for which all
three of these phases is stable is 2:1 lauric acid (LA)/dilauroyl
phosphatidylcholine (DLPC) and water (with P coexisting with excess
water) \cite{a:temp98a}. The only other system for which G, D and P
are known to be stable is didodecylphosphatidylethanolamine and water
\cite{sedd90}; however, here G exists with hardly no hydration at all,
and the regions for D and P are very difficult to access
experimentally, thus this system should not be compared with the
theoretical results for fully hydrated phases presented here.  In all
other systems with IBCPs, only one or two of them are stable (for
example, for monoolein and water, G and D are stable, with D
coexisting with excess water
\cite{a:hyde84}).  In this work we present a model which can explain
why only G, D and P are observed in lipid-water mixtures and why their
actual occurrence seems to depend on the specific system under
investigation.  The phase diagram predicted by our model agrees
well with the experimental phase diagram for 2:1 LA/DLPC and water.

Exact (\emph{Weierstrass}) representations are only known for G, D, P
and I-WP \cite{a:lidi90,a:fodg92b,a:cvij94,a:fodg99}.
However, it has been shown in recent studies
\cite{a:gozd96b,uss:schw99} that these and other cubic TPMS can be
generated as isosurfaces of a density field, which minimizes the
free-energy functional of a simple Ginzburg-Landau model for
amphiphilic systems \cite{a:gomp90}.  Using the Fourier
representations of Ref.~\cite{uss:schw99}, it is now possible to
investigate also the physical properties of the IBCPs, for whose
minimal mid-surfaces no exact representations are known. In
particular, the question of whether these structures might correspond
to stable IBCPs in lipid-water mixtures can now be addressed
quantitatively.  In this work we study G, D, P, I-WP, S, C(P) and
F-RD, of which G, D, P, S and C(P) are balanced.

The most important contribution to the free energy of amphiphilic
interfaces is the curvature energy, which dominates all other
contributions if the radii of curvature are large compared with
molecular length scales \cite{a:safr99}. For the inverse phases in
lipid-water mixtures, it is generally assumed that another important
contribution to the free energy is the stretching energy of the
hydrocarbon chains
\cite{a:ande88,a:dan93,a:dues97}. In order to attain optimal free
energy, the neutral surfaces of the two monolayers should realize at
the same time constant mean curvature $c_0$ and constant distance $l$
to the minimal mid-surface.  Since this is not possible geometrically,
the free energy will always be frustrated \cite{a:char90}.  The
relative importance of interface (bending) and bulk (stretching)
contributions in self-assembled interfaces is a long-debated issue.
For amphiphilic systems, for which the free energy contributions come
mainly from the interfaces, the main problem is of a mathematical
nature: the overall solution can be investigated only in a model which
allows the interfacial positions to adjust themselves freely.  In
contrast, in diblock copolymer systems bulk contributions are much
more important and models based on random polymer coils and
interacting monomer densities are more adequate.  Self-consistent
mean-field theory for diblock-copolymer systems suggests that in this
case, interfacial tension and stretching are equally important
\cite{a:mats96b}. Recently, this framework has been adapted to treat
lipid-water mixtures \cite{a:li00}.  This approach is complementary to
the membrane approach; it is most useful for the investigation of
systems, in which the internal structure of the bilayer is not
uniform.  However, self-consistent field theory for large molecules is
not expected to describe amphiphilic systems well at large lattice
constants, when the membranes' internal structure is essentially
preserved and the main contributions to the free energy depends on
their shape. In particular, only in the membrane approach does it
becomes possible to address the role of thermally activated membrane
undulations.

In order to study interface-dominated systems, it has been suggested
in Ref.~\cite{a:ande88} to consider the two classes of
surfaces, which can completely relax one of the two relevant
contributions. {\em Parallel surfaces} to the minimal midsurface
realize the optimal chain length $l$ and are frustrated only in
their bending energy, while {\em constant-mean-curvature surfaces}
(CMC-surfaces) realize the given spontaneous curvature, but are
frustrated in their stretching energy.  The mathematical properties of
parallel surfaces are well known in terms of the properties of the
underlying minimal surface; the parallel-surface model has therefore
been the main tool for the investigation of IBCPs
\cite{a:ande88,a:hyde89,a:helf90,a:temp98a}.  Up to now, the main
difficulty of this approach has been to take into account the
variation of Gaussian curvature over the minimal mid-surface.  In
earlier work, the curvature energy was therefore expanded in moments
of the distribution of the Gaussian curvature.  In
Ref.~\cite{a:ande88} this was done up to eighth order for D, and in
Ref.~\cite{a:helf90} up to second order for P, D and G.  CMC-surfaces
\cite{a:ande88,a:dues97,a:temp98a} have the advantage that they offer
an unified picture for all relevant phases --- lamellar, inverse
bicontinuous cubic, inverse hexagonal and inverse micellar cubic
phases are modelled by planes, cubic CMC-surfaces, cylinders and
spheres, respectively. However, since no exact representations are
known, the stretching frustration for IBCPs has so far been evaluated
numerically for D only \cite{a:ande88}.

We will show below that in the case of short chains (or large lattice
constants), stretching energy is prohibitively large, and the two
monolayers of a IBCP in lipid-water systems can be modeled by parallel
surfaces.  In the case of long chains (or small lattice constants), we
expect stretching and bending to be equally important; this agrees
with the results for diblock copolymers, which are in the long chain
limit. Our reasoning implies that CMC-surfaces are not a good
representation of IBCPs in lipid-water systems for physical
reasons. However, for small distances to the minimal mid-surface,
their mathematical properties are very similar to the ones of parallel
surfaces \cite{a:ande88,uss:schw00a}, thus many results from the
parallel surface model will carry over to the CMC-model. In general,
however, the parallel surface model is more appropriate both for
physical and mathematical reasons. Moreover, there is strong
experimental evidence for it, both from a detailed reconstruction of
electron densities \cite{a:luzz93} and from analysis of swelling data
\cite{a:engb95}.

In contrast to earlier work, we evaluate the curvature energy in the
parallel surface model without any expansion and to high numerical
accuracy by first calculating the distribution of Gaussian curvature
over the minimal mid-surface.  For G, D, P and I-WP this can be done
from their Weierstrass-representations.  We also consider the
structures S, C(P) and F-RD by calculating the distribution from the
representations which we obtained from Ginzburg-Landau theory. Our
main result is that the free-energy densities of the structures G, D
and P are considerably lower than those of the other investigated
structures due to their narrow distribution of Gaussian curvature over
the minimal midsurface. We show that this result persists when thermal
membrane undulations and higher order terms in the bending energy are
considered, and argue that potential additional contributions to the
free energy are unlikely to change it.  This explains why only G, D
and P have been observed in lipid-water mixtures. In fact, the
calculated phase diagram agrees nicely with the experimental one for
2:1 LA/DLPC and water, for which these three phases coexist. We show
that due to the existence of a Bonnet transformation between G, D and
P, these phases coexist along a triple line in our model.
Simultaneously, P coexists with an excess excess water due to a
mechanism called {\em emulsification failure}.  The Bonnet
transformation also implies that the free-energy densities of G, D and
P scale as a function of concentration with a universal geometrical
quantity, which we term \emph{topology index}.  Since the topology
index decreases from G to D to P, the gyroid G is most prominent,
followed by smaller regions of stability for D and P at higher water
concentrations. Any additional contribution to the free energy, which
introduces a new length scale, is expected to change the delicate
balance between these three phases; this includes stretching
contributions, van der Waals and electrostatic interactions.  Since
such additional effects are specific to a given experimental system,
this result explains qualitatively why usually only one or two of
these phases are observed in lipid-water mixtures.

The paper proceeds as follows. In Sec.~\ref{sec:Model}, we specify the
free-energy expression which has to be evaluated for each type of
minimal mid-surface, and discuss some immediate consequences of the
model.  The calculation of the distribution of Gaussian curvature is
explained in Sec.~\ref{sec:Weierstrass} for P, D, G and I-WP from
their Weierstrass representation. Results for the Gaussian-curvature
distributions for S, C(P) and F-RD are obtained in
Sec.~\ref{sec:GinzburgLandau} from the representations derived
recently from a simple Ginzburg-Landau model. These results are
combined in Sec.~\ref{sec:PhaseDiagrams} to predict phase behavior and
to numerically calculate phase diagrams. In
Sec.~\ref{sec:symmetry_breaking}, we discuss which physical mechanisms
will break the Bonnet symmetry between P, D and G.  Finally, we
compare our results with experiments and conclude in
Sec.~\ref{sec:Discussion}.

\section{The model}
\label{sec:Model}

The bending energy of one lipid monolayer is described by the
\emph{Canham-Helfrich Hamiltonian}
\cite{a:canh70,a:helf73}
\begin{equation} \label{HelfrichHamiltonian}
E_b = \int dA^l\ \left\{ 2 \kappa\ (H^l - c_0)^2 + \bar \kappa\ K^l \right\}\ ,
\end{equation}
where the integration extends over the monolayer's neutral surface,
which is at a distance $l$ from the minimal midsurface of the bilayer.
In the framework of the curvature model, bending and stretching
contributions decouple at the neutral surface \cite{a:safr99};
this allows to disregard area-stretching contributions. 
The neutral surface has been shown experimentally to be located close 
to the polar-apolar interface \cite{temp95}. 
In the curvature model (\ref{HelfrichHamiltonian}),
the neutral surface is characterized by mean curvature $H$, Gaussian
curvature $K$ and differential area element $dA$; the superscript $l$
indicates that the neutral surface is a parallel surface.  The model
parameters are the spontaneous curvature $c_0$, the bending rigidity
$\kappa$ and the saddle-splay modulus $\bar \kappa$ of the monolayers.
For lipids, the bending rigidity $\kappa$ for a monolayer is of the
order $10-20\ k_B T$. Usually the saddle-splay modulus $\bar \kappa$
is assumed to have a small negative value; in fact the curvature model
\eq{HelfrichHamiltonian} without any additional constraints is
well-defined only for $-2 \kappa \le \bar \kappa \le 0$
\cite{a:helf81}.  The spontaneous curvature $c_0$ can be considered to
result from the mismatch between optimal head group area, volume
incompressibility of the chains and optimal chain length. For
non-ionic lipids, it depends mainly on temperature, which changes the
head group hydration and therefore optimal head group area.  Since
lipids typically form lamellar and inverse phases, where the
monolayers prefer to bend towards their polar sides, we define
positive curvature to be towards the water regions.  We assume that
spontaneous curvature scales with temperature as $c_0 \sim (T - T_b)$,
since such a linear relationship seems to hold very well for ternary
surfactant systems \cite{a:stre94}.

The stretching energy can be assumed to be harmonic about the average 
chain length $l$, so that
\begin{equation} \label{eq:stretching}
E_s = \int dA\ k_s\ (L-l)^2 
\end{equation}
where $L$ is the local chain length.  The stretching modulus $k_s$ is
itself a function of the average chain length $l$. The relevant
scaling law has to be obtained from a microscopic model; here, we model
the lipid monolayer as a Gaussian polymer brush grafted to the
polar-apolar interface, but more realistic models  
give similar results \cite{a:safr99}.  The stretching energy per chain
reads $E_s = (3 k_B T/2 N) (l/a_K)^2$, where $a_K$ is the
(microscopic) Kuhn length and $N$ the number of Kuhn segments.
Enforcing the volume constraint $l d^2 = N a_K^3$ (where $1/d^2$ is
the grafting density), and transforming from energy per chain to
energy per area, results in a potential of the form \eq{eq:stretching}
with $k_s \sim k_B T (a_K / l d^4)$.  Thus the stretching modulus
$k_s$ scales inversely with average chain length $l$ --- stretching
becomes very difficult for short chains \cite{a:safr99}.

Although the bending energy \eq{HelfrichHamiltonian} is well known not
to depend on microscopic details, for our purpose it is important to
note that the bending rigidity $\kappa$ also depends on
the average chain length $l$. Starting from the stretching 
energy of \eq{eq:stretching} and taking into account how the 
chain volume varies with curvature, results in the bending energy of 
\eq{HelfrichHamiltonian} with $\kappa \sim k_s l^4 \sim k_B T (a_K l^3/ d^4)$
\cite{a:safr99}. The result that the bending rigidity $\kappa$
scales with the third power of film thickness $l$ is well known also
from the elasticity of thin solid sheets.

We now can estimate the relative importance of bending and stretching
contributions. It is known that for both CMC-surfaces and parallel
surfaces, the normalized standard deviations of $L$ and $H$,
respectively, are almost independent of hydrocarbon volume fraction
\cite{a:ande88}, so that $\langle (L-l)^2 \rangle \sim l^2$ and $\langle (H -
\langle H \rangle)^2\rangle \sim \langle H \rangle^2 \sim a^{-2}$, where $a$
is the lattice constant. Therefore $E_s / E_b \sim k_s l^2 / \kappa
a^{-2} \sim (a / l)^2$.  Thus, for small chain length $l$ (large
lattice constant $a$), the stretching term is much larger and the
chains can be assumed to be of constant length. This corresponds to
the well-known result that the curvature energy is the relevant
contribution to the free energy as long as the curvature radii (that
is the lattice constants) are much larger than any molecular length
\cite{a:safr99}.  Only for large chain length $l$ (small lattice
constant $a$), the bending energy becomes comparable and both terms
have to be considered.  Experimentally, this corresponds to the fact
that for small water content (i.e. chain length $l$ of comparable size
as lattice constant $a$), the bicontinuous cubic phases become
unstable with respect to the hexagonal phase, whose geometrical
properties require larger chain stretching.  Note that the relative
importance of stretching and bending is asymmetric: since the chain
length $l$ cannot be larger than the lattice constant $a$, there is no
regime in which bending can be neglected.  Thus CMC-surface will never
be a good approximation for monolayers in lipid-water systems.
However, the stretching frustration can be relaxed by swelling the
bilayer with oil; therefore CMC-surfaces are good approximations for
amphiphilic monolayers in ternary systems with oil, water and
amphiphile \cite{uss:schw00a}.  Similar considerations of the relative
importance of stretching and bending can be found also in the context
of large membrane inclusions (like transmembrane proteins) where
chains have to stretch in order to decrease the hydrophobic mismatch
at the inclusion boundary \cite{a:dan93}.

For the rest of this paper, we proceed within the framework of the
parallel surface model.  We assume that the interfaces between the
tail and head group regions of the two lipid monolayers in a ICBP are
located at distances $\pm l$ away from a given cubic TPMS; typically
$l$ has values between one and two nanometers.  Note that for a TPMS
the mean curvature satisfies $H = (c_1 + c_2)/2 = 0$ everywhere (where
$c_1$ and $c_2$ are the two principal curvatures), while the Gaussian
curvature $K = c_1 c_2 = - {c_1}^2$ varies as function of the position
and is only restricted to satisfy $K \le 0$.  The differential area
element $dA^l$, the mean curvature $H^l$ and the Gaussian curvature
$K^l$ on the parallel surfaces then follow as functions of $l$ and the
quantities $dA$ and $K$ on the minimal mid-surface
\cite{a:spiv79,a:hyde89}, with
\begin{equation} \label{eq:ParallelSurfaces}
dA^l = dA\ (1 + K l^2)\ , \quad
H^l  = \frac{- K l}{1 + K l^2}\ , \quad
K^l  = \frac{K}{1 + K l^2}\ .
\end{equation}
Note that these formulae are special cases of Steiner's theorem of integral
geometry; they are therefore no approximations for small $l$, but
exact.  Since minimal surfaces have $K \le 0$, positive mean curvature
is defined here to correspond to positive $l$.

The effective bending energy of the lipid bilayer is only a function of the
Gaussian curvature $K$ of the minimal midsurface since its mean
curvature $H$ vanishes. In an expansion for small chain length $l$, it
can be expressed as an infinite series in powers of $K$. Using
\eq{HelfrichHamiltonian} and \eq{eq:ParallelSurfaces} yields up to order
$l^2$
\begin{equation}
  \label{eq:EffectiveBilayerBending}
  E_b = \int dA \left\{ 4 c_0^2 l^2 \kappa 
         + (2 \bar \kappa + 8 c_0 l \kappa + 4 c_0^2 l^2 \kappa) K 
         + 4 \kappa l^2 K^2 \right\}\ .
\end{equation}
Therefore the effective saddle-splay modulus for the lipid bilayer,
$\bar{\kappa}_{bi}= 2 \bar \kappa + 8 c_0 l \kappa + 4 c_0^2 l^2
\kappa$, is not just the sum of the monolayer's moduli, but is
corrected to higher positive values due to the presence of the
spontaneous curvature $c_0$ \cite{a:port92}.
We conclude that as long as $c_0 l \gtrsim -
\bar{\kappa} / 4 \kappa$, the IBCPs are favored over the lamellar
phase, since the preferred curvature of the monolayers translates into
a topological advantage of saddle-type bilayer structures. The third
term in \eq{eq:EffectiveBilayerBending} favors saddle-type structures
in any case, and also removes the degeneracy which minimal surfaces
might experience under deformations which preserve the minimal surface
property.  This has been noted in a seminal work by Bruinsma
\cite{a:Brui92}; in his notation, we have $\bar{\bar{\kappa}} = 4
\kappa l^2$ for the elastic modulus of the $K^2$-term.

The volume fraction occupied by the hydrocarbon can be calculated as
\begin{equation} \label{hydrocarbonvolume}
v = \frac{1}{a^3} \int_{- l}^{l} d{l'} \int dA^{l'} 
  = 2 A^* \left( \frac{l}{a} \right) 
                 + \frac{4 \pi}{3} \chi \left( \frac{l}{a} \right)^3     
\end{equation}
where we have used the Gauss-Bonnet theorem $\int dA K = 2 \pi \chi$.
$A^*$ and $\chi$ denote surface area and Euler characteristic of the
given cubic TPMS in the conventional unit cell with unit lattice
constant.  For the following, it is useful to introduce the
\emph{topology index} $\Gamma = ({A^*}^3 / 2 \pi |\chi|)^{1/2}$.  This
quantity is independent of scaling and choice of unit cell and
characterizes a given TPMS's topology in an universal way; the higher
its value, the smaller is the porosity and the larger the specific
surface area of a TPMS. It is the only quantity which characterizes a
two-dimensional minimal surface in three-dimensional space
independently of lattice constant and choice of unit cell, and its
relevance for the description of bicontinuous cubic phases in
amphiphilic systems has been discussed before
\cite{a:hyde89,a:stro92,uss:schw99,uss:schw00a,a:fodg99}. The topology
index of the various TPMS is of order $1$ and is given in
Tab.~\ref{TableIndex}.  The gyroid G is expected to have the highest
value since it divides space into labyrinths with threefold
coordinated vertices.  S has a similar value for $\Gamma$, since it
features a mixture of three- and fourfold coordinated vertices.  Then
comes the double diamond structure with fourfold coordinated vertices,
while all other structures have higher values. In
Tab.~\ref{TableIndex} we collect the values of $A^*$, $\chi$ and
$\Gamma$ for the cubic TPMS investigated in this work as obtained from
our numerical representations in the framework of a simple
Ginzburg-Landau theory (see below). Exact values are known for G, D,
I-WP, P and C(P) and given in Tab.~\ref{TableNumbers}.  Comparing
numerical and exact values shows that for most structures, our
numerical results are quite accurate.  Due to numerical limitations
for the more complicated structures, the values for S and F-RD are
less precise.  In particular, we expect S to have a smaller value for
$\Gamma$ than G.

\begin{figure}
\begin{center}
\leavevmode 
\psfig{file=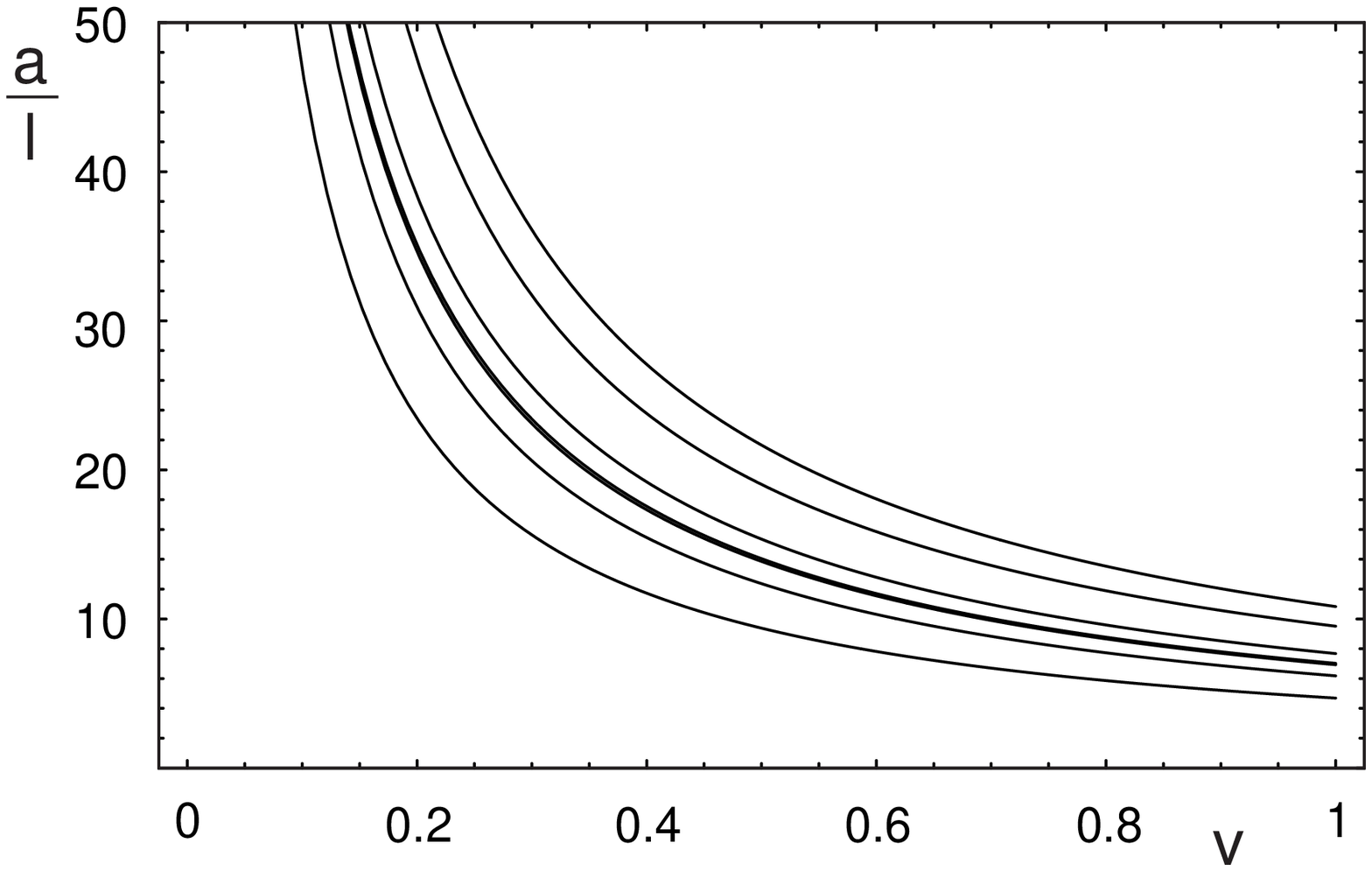,width=0.8\textwidth}
\end{center}
\caption{Lattice constant $a$ in units of the chain length $l$ 
  as a function of hydrocarbon volume $v$. We use $a/l = v/2 A^*$,
  which is an excellent approximation up to $v \approx 0.8$, beyond
  which the inverse bicontinuous cubic phases as modeled here cannot
  exist anymore due to self-intersection. From left to right, the
  curves correspond to P, G, I-WP, C(P), D, F-RD and S, respectively
  (the curves for I-WP and C(P) nearly collapse and therefore appear
  as an apparent bold line).}
\label{FigureLatticeConstant}
\end{figure}

Equation (\ref{hydrocarbonvolume}) can be inverted numerically to
give the dimensionless function $a/l$, the lattice constant $a$ in
units of the chain length $l$, as a function of hydrocarbon volume
$v$. For small $v$ we find
\begin{equation} \label{LatticeConstant}
\frac{a}{l} = \frac{2 A^*}{v} \left( 1 + \frac{1}{12\ \Gamma^2} v^2 
+ {\cal O}(v^4) \right)\ .
\end{equation}
In Fig.~\ref{FigureLatticeConstant} we plot the first-order
approximation $a/l = 2 A^* / v$ of \eq{LatticeConstant} for all
structures considered; it can be hardly distinguished from the full
curve over the full range of $v$. By visual inspection of our
numerical representations (see below), we find that for all structures
considered, parallel surfaces do not self-intersect as long as $v
\lesssim 0.8$; IBCPs can accommodate large amounts of hydrocarbon
since their geometry is so close to that of the lamellar phase. A
typical value for $l$ is $15 \AA$, so that for $v = 0.5$ we have
$a_{S} = 325\ \AA$, $a_{F-RD} = 285\ \AA$, $a_D = 230\ \AA$, $a_{CP} =
211\ \AA$, $a_{I-WP} = 208\ \AA$, $a_G = 185\ \AA$ and $a_P = 141\
\AA$.  Note that at a given $v$, the ratio of the lattice constants of
two coexisting IBCPs is simply the ratio between their scaled surface
areas $A^*$. In Figs.~\ref{FigureParallelSurfaces1} and
\ref{FigureParallelSurfaces2} we depict for each of the structures
considered one of the two monolayers for the hydrocarbon volume
fraction $v = 0.5$. Then the volume fraction of each of the two
labyrinths is $(1-v)/2 = 0.25$.

We are now in a situation to rewrite the curvature energy from
\eq{HelfrichHamiltonian} for our purpose. We consider a typical
experimental situation, in which temperature is controlled at constant 
volume. As explained above, changing temperature $T$ amounts to
changing spontaneous curvature $c_0$. Under the assumption of 
incompressibility of
both components, there is only one degree of freedom for composition,
which we take to be the hydrocarbon volume $v$ (which is equivalent to
the amount of interfacial area).  Dividing the free energy by the
constant volume, we arrive at the free energy per {\em unit volume} as
a function of $v$. However, $v$ is not a controlled quantity, since at
two phase coexistence, it is not clear how it will distribute into the
two phases. The controlled quantity can be considered to be the
chemical potential coupled to $v$ (that is the derivative of free
energy -- or free energy volume density -- for hydrocarbon volume
fraction -- or interfacial area), and therefore phase transitions have
to be calculated by using the Maxwell construction for the free energy
density as a function of $v$ (which amounts to requiring the same
chemical potential for both phases).  We want to remark
parenthetically that the free energy per {\em unit area} cannot be
used to calculate phase coexistences, since an exchange of area,
controlled by the chemical potential of the lipid, has to be allowed
between coexisting phases.
 
We write the free energy volume density dimensionless by using a factor
$l/(2\kappa c_0^2)$.  For the lamellar phase the hydrocarbon volume
fraction $v = 2 l/a$, thus its dimensionless free-energy density
follows from
\eq{HelfrichHamiltonian} to be 
\begin{equation} \label{FreeEnergyDensityLamellar}
f_L = \frac{l}{2 \kappa {c_0}^2} \frac{1}{A a}  4 \kappa c_0^2 A = v\ . 
\end{equation}
The larger $v$, the smaller $a$ and the more frustrated curvature
energy per volume accumulates. 

For IBCPs, we again use \eq{HelfrichHamiltonian} and
\eq{eq:ParallelSurfaces} in order to write the dimensionless 
free-energy density as a function of hydrocarbon volume fraction $v$ and
the properties of the bilayers' minimal mid-surface:
\begin{equation} \label{FreeEnergyDensityBicont}
f_b = 2 \left( \frac{l}{a} \right) \int dA^* 
               \left( 1 + K^* \left( \frac{l}{a} \right)^2 \right) 
   \left( \frac{K^* \left( \frac{l}{a} \right)^2}
          {(1 + K^* \left( \frac{l}{a} \right)^2) c_0 l} + 1 \right)^2 
        - r \frac{4 \pi \chi}{c_0^2 l^2} \left( \frac{l}{a} \right)^3
\end{equation}
where again we have used the Gauss-Bonnet theorem and defined $r = -
\bar \kappa / 2 \kappa$.  As mentioned above, the curvature model
\eq{HelfrichHamiltonian} without any additional constraints is
well-defined for $0 \le r \le 1$ and bicontinuous phases are favored
for $c_0 l \gtrsim r/2$.  $A^*$, $K^*$ and $\chi$ refer to a
conventional unit cell.  Using \eq{LatticeConstant} to lowest order,
$a/l = 2 A^* / v$, and defining $\Xi(K^*) = K^* A^* / 8 \pi \chi$, we
rewrite \eq{FreeEnergyDensityBicont} as
\begin{eqnarray} \label{FreeEnergyDensityBicont2}
f_b = v\ \Big\{ \int \frac{dA^*}{A^*}
\left( 1 -  \Xi(K^*) \left(\frac{v}{\Gamma} \right)^2 \right)^{-1} 
\left( 1 - \frac{1+c_0}{c_0} \Xi(K^*) \left(\frac{v}{\Gamma} \right)^2 \right)^2
+ \frac{r}{4 c_0^2} \left( \frac{v}{\Gamma} \right)^2 \Big\}\ .
\end{eqnarray} 
In Eq.~(\ref{FreeEnergyDensityBicont2}) and below, spontaneous
curvature $c_0$ is measured in units of $1/l$.  Since for the minimal
midsurface both $K^*$ and $\chi$ are negative, $\Xi(K^*)$ is a
positive quantity which varies over the surface.  Note that $\int
(dA^* / A^*) \Xi(K^*) = 1/4$ due to the Gauss-Bonnet theorem. The
free-energy density $f_L$ of the lamellar phase,
Eq.~(\ref{FreeEnergyDensityLamellar}), is a special case of $f_b$,
with $\Xi(K^*) = \chi = 0$.  In this work, we will treat the
free-energy density \eq{FreeEnergyDensityBicont2} without any further
approximations.  Previously, similar expressions have been expanded in
small $v$ (which is equivalent to small $l$) \cite{a:ande88,a:helf90},
similarly as discussed in \eq{eq:EffectiveBilayerBending}. In fact,
each term $(K l^2)^n$ in an expansion for small $l$ corresponds to a
term $v^{2n+1}$ in an expansion for small hydrocarbon volume $v$.
Since $K l^2 = - \Xi(K^*) (v/\Gamma)^2$, higher-order terms in
\eq{HelfrichHamiltonian} would result in a free-energy density which
has a similar structure as \eq{FreeEnergyDensityBicont}. As we will
see below, this means that our results will remain valid in the case
that higher-order terms are included in the curvature energy of the
monolayers.

It is well known that thermal fluctuations can contribute
significantly to the free-energy of amphiphilic systems. Since we
consider the regime where the stretching energy dominates, undulations
are the favored fluctuation modes. For the lamellar phase,
they give rise to steric repulsion \cite{a:helf78}, which in our case can be
written as 
\begin{equation} \label{eq:LamellarSteric}
f_{steric} = \frac{c_\infty}{32 c_0^2} \
\left(\frac{k_BT} {\kappa}\right)^2 \ \frac{v^3}{(1 - v)^2}
\end{equation}    
where $c_\infty = 0.106$ from Monte Carlo simulations 
\cite{a:gomp89b,a:jank89a} and field-theoretical calculations 
\cite{a:bach99}. For an IBCP, steric repulsion is hardly relevant since 
the lateral correlation length $\xi_\parallel$ is of the order of the
lattice constant $a$.  Therefore, the perpendicular correlation length
$\xi_{\perp} \sim \sqrt{k_B T / \kappa}\ \xi_{\parallel} \ll
\xi_{\parallel} \sim a$, and few membrane collisions should occur.
However, fluctuations of the lipid bilayer lead to a renormalization
of $\kappa$ and $\bar \kappa$ \cite{a:nels89} for all phases.  Since
$\kappa$ multiplies the average mean curvature squared, which vanishes
for the mid-plane, only the renormalization of $\bar\kappa$ at length
scale $\ell$ has to be taken into account.  We identify the typical
length scale of a cubic structure with $\langle K \rangle^{-1/2}$.
This implies $\ell/l = 2 \Gamma / v$, so that $r$ in
Eq.~(\ref{FreeEnergyDensityBicont2}) gets renormalized, with
\begin{equation} \label{renormalization}
r_R = \left[ r - \frac{5}{12\pi} \frac{k_BT}{\kappa}
 \ln\left(\frac{2 \Gamma}{v} \right) \right]\ .
\end{equation}                                                                      
Since the renormalization increases $\bar \kappa$, this effect
increases the topological advantage of the saddle-shaped structures as
does the spontaneous curvature. Note that for the lamellar phase,
these effects might favor the formation of wormholes \cite{a:Gomp95c}.
Here, we neglect this aspect, as well as the acoustic modes of the
bicontinuous structures at large wavelengths, which have been
discussed by Bruinsma \cite{a:Brui92}. Note also that the thermal
fluctuations also give a contribution to the free energy, which is
linear in the membrane area. We have omitted this term here, since it
can be absorbed into the chemical potential of the lipid. 

Having defined the model, we now turn to its phase behavior.  We
first note that the dimensionless mean curvature averaged over the
parallel surface follows from \eq{eq:ParallelSurfaces} as
\begin{equation} \label{eq:GeometricalProperty}
\langle H^l \rangle_l l = \frac{\int dA^l\ H^l l}{\int dA^l}
= \frac{(v / \Gamma)^2}{4 - (v / \Gamma)^2}
\end{equation}
where again $l/a = v / 2 A^*$ and the Gauss-Bonnet theorem has been
used.  It is plotted in Fig.~\ref{FigureGeometry1}a.  Note that all
curves fall on a universal curve when the hydrocarbon volume fraction
$v$ is scaled with the topology index $\Gamma$. It reaches the value
$c_0$ at
\begin{equation} \label{eq:MeanCurvatureAtc0}
v = \left( \frac{4 c_0}{1 + c_0} \right)^{\frac{1}{2}}\ \Gamma\ .
\end{equation}
Thus, $c_0$ is realized in a sequence given by increasing topology
index; in particular, the gyroid structure G reaches $c_0$ at the
highest value of $v$. We conclude that if all considered structures
were stable, they would appear in the sequence G - S - D-
I-WP - P etc with increasing water concentration.

The stability of the different phases is determined by the free-energy
density given in \eq{FreeEnergyDensityBicont}.  The second
(topological) term is easy to understand. The topological properties
of the parallel and mid-surfaces are the same: for negative
saddle-splay modulus $\bar \kappa$ ($r > 0$), the different IBCPs are
favored according to their values of the topology index $\Gamma$.
Here the G-structure performs best since it has the lowest porosity.
The first term in \eq{FreeEnergyDensityBicont} is more complicated,
since the mean-curvature properties of the parallel surfaces translate
into the Gaussian-curvature properties of the minimal mid-surface in a
quite complicated way.  As discussed above, the spontaneous curvature of
the monolayers implies a topological advantage of the
bicontinuous phases.  For a more detailed analysis, which includes all
terms of the expansion, we note that the first term in
\eq{FreeEnergyDensityBicont} measures the standard deviation of the
parallel surface's mean curvature from the spontaneous curvature.
Since it follows from \eq{eq:ParallelSurfaces} that the distribution of
mean curvature over the parallel surface, $H^l dA^l$, is proportional
to the distribution of Gaussian curvature over the minimal surface, $K
dA$, we expect that those structures will be more favorable which have
small standard deviations for their Gaussian curvature distributions
\cite{a:helf90}. In order to quantify this concept, we now turn to the
distributions of Gaussian curvature which later will allow us to
evaluate \eq{FreeEnergyDensityBicont} without any further
approximations.

\section{Curvature properties from Weierstrass Representations}
\label{sec:Weierstrass}

\begin{figure}
\begin{center}
\leavevmode 
\psfig{file=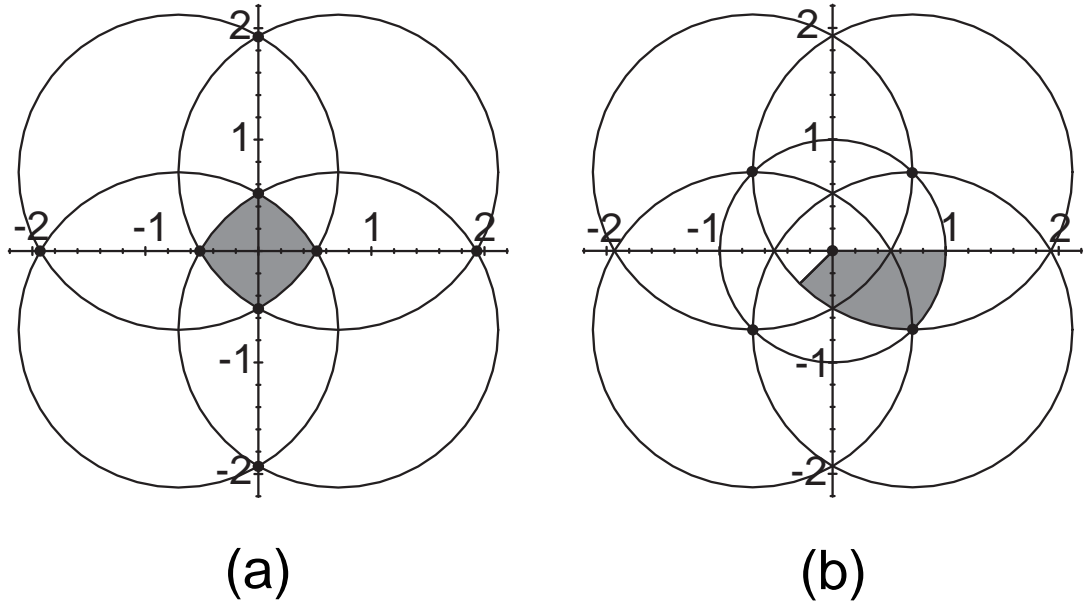,width=0.8\textwidth}
\end{center}
\caption{Fundamental domains for Weierstrass representations: (a) G, D and P
  and (b) I-WP. Filled circles mark the poles of the generating
  functions $R(z)$ (which correspond to the flat points of the surfaces)
  and the hatched regions correspond to the fundamental domains of the
  TPMS.}
\label{FigureWeierstrass}   
\end{figure}

Weierstrass representations are known for P, D and G
\cite{a:fodg92b,a:fodg99} and I-WP \cite{a:lidi90,a:cvij94}.
For each of these TPMS, a fundamental domain can be identified, so that
the rest of the surface follows by replicating it with the appropriate
space group symmetries ($Im\bar{3}m$, $Pn\bar{3}m$, $Ia\bar3d$ and
$Im\bar{3}m$, respectively). The Weierstrass representation is a
conformal mapping of certain complicated regions within the complex
plane onto the fundamental domain:
\begin{equation} \label{WeierstrassDarstellung}
(x_1,x_2,x_3) = Re \int_0^{u + i v} dz\ R(z)\ (1 - {z}^2,i (1 + {z}^2),2 z)
\end{equation}         
where $(u,v)$ are the internal (and conformal) coordinates of the
minimal surface. The geometrical properties of such a surface follow as
\begin{equation} \label{WeierstrassDarstellung2}
dA(z) = |R(z)|^2\ (1 + |z|^2)^2\ du dv,\ \quad
H(z) = 0,\ \quad
K(z) = \frac{-4}{|R(z)|^2 (1 + |z|^2)^4}
\end{equation}    
with $z = u + i v$. Obviously the (isolated) poles of $R(z)$
correspond to the flat points ($K = 0$) of the minimal surface. Only
few choices of $R(z)$ yield embedded minimal surfaces.  The ones for D
and P have been known since the 19th century from the work of Schwarz:
for D it is $R(z) = (z^8 - 14 z^4 + 1)^{-\frac{1}{2}}$.  P corresponds
to the same region and follows simply by the Bonnet transformation
$R(z) \to e^{i \theta} R(z)$ with $\theta = 90^o$.  Equation
(\ref{WeierstrassDarstellung2}) implies that P and D have the same
metric and the same distribution of Gaussian curvature.  However,
since they map differently into embedding space, they have different space
groups and lattice constants. The gyroid G was discovered in 1970 by
Schoen \cite{a:scho70} as another Bonnet transformation of D, with
$\theta = 38.015^o$. The Weierstrass representation for I-WP was found
only recently \cite{a:lidi90,a:cvij94}. If one of its poles is chosen
to be at infinity, one has $R(z) = (z (z^4 + 1))^{-\frac{2}{3}}$.  In
Fig.~\ref{FigureWeierstrass} we show the fundamental domains in the
complex plane of the different Weierstrass representations.  In
Tab.~\ref{TableNumbers} we collect the values for scaled surface area
$A^*$ and lattice constant $a$ as they follow from the Weierstrass
representations. We also give the number $N$ of replications of the
fundamental domain needed to build up the surface in one conventional
unit cell and the values for the Euler characteristic $\chi$ which
follows from the topology generated in this process. Note that for D
one can choose another conventional unit cell which is contained $n =
8$ times in the one chosen here.  This has been done in
Refs.~\cite{a:ande88,a:temp98a}; then $\chi$, $A$ and $K$ scale as
$n^{-1}$, $n^{-1/3}$ and $n^{-2/3}$, respectively. In particular, one
then has $\chi = -2$ and $A = 1.9188925$. Our choice is motivated by
the fact that it is the relevant one also for ternary systems. 

We define the distribution function of Gaussian curvature,
\begin{equation} \label{eq:DistributionFunction}
  f(K) = \int dA(u,v)\ \delta(K - K(u,v))  \ ,
\end{equation}
over some surface parametrized by internal coordinates $(u,v)$,
where in our case $dA(u,v)$ and $K(u,v)$ are determined by 
\eq{WeierstrassDarstellung2}. For the generating functions $R(u,v)$
detailed above, it is not possible to calculate $f$ analytically.
Therefore, we evaluate it numerically by considering a histogram $\{
f_i \}$ for a set of discrete Gaussian curvature values $\{ K_i \}$ spaced
equidistantly with $\Delta K$. We cover the $(u,v)$-region
corresponding to the fundamental domain with a square grid of $M
\times M$ points. For each point, we calculate $dA$ and $K$ from
\eq{WeierstrassDarstellung2} and add $dA$ to the $f_i$ which
corresponds to the $K_i$ with $K \in [K_i - \Delta K / 2, K_i + \Delta K /
2]$. In order to obtain results for one conventional unit cell, the
values $\{ f_i \}$ are then multiplied by the number $N$ of
replications and the values $\{ K_i \}$ are scaled with 
$a^2$ (both $N$ and $a$ are given in
Tab.~\ref{TableNumbers}).  In practice we use $M = 30000$ and $\Delta
K = 0.025$; the latter value amounts to about $150$ bins for the
histogram. By this procedure, we have
\begin{equation} \label{eq:DistributionFunctionNumerical}
f_i = \int dA(u,v)\ \int_{K_i - \Delta K / 2}^{K_i + \Delta K / 2}
dK'\ \delta(K' - K(u,v)) = \Delta K f(K_i)\ .
\end{equation}
The moments $S_N$ of the distribution $f(K)$ follow as
\begin{equation} \label{eq:DistributionFunctionMoments}
  S_N = \int dK f(K) K^N = \int dA K^N = \sum_i f_i {K_i}^N\ .
\end{equation}
In particular, $S_0 = \sum_i f_i = A^*$ and $S_1 = \sum_i f_i K_i = 2 \pi
\chi$. 

\begin{figure}
\begin{center}
\leavevmode 
\psfig{file=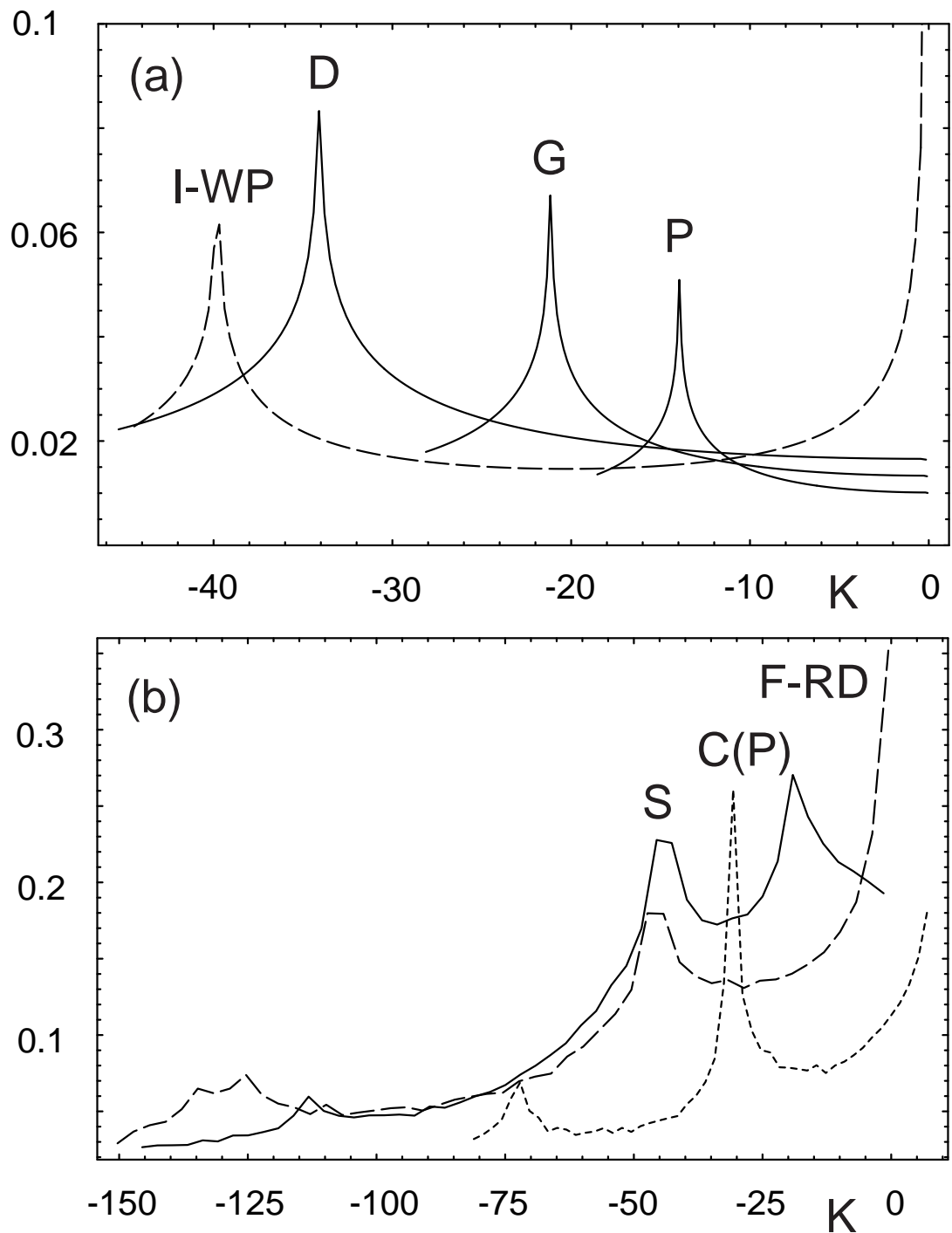,width=0.8\textwidth}
\end{center}
\caption{Distributions of Gaussian curvature for (a) 
  I-WP, D, G and P as obtained from the Weierstrass representations
  and (b) S, C(P) and F-RD as obtained from our numerical
  representations as isosurfaces to Ginzburg-Landau local minima.
  Since P, D and G are related by a Bonnet-transformation, they differ
  only by simple rescaling. This data has also been shown in 
  Ref.~\protect\cite{uss:schw99}.}
\label{FigureDistributions1}
\end{figure}

The procedure described here was carried out to obtain $\{K_i, f_i\}$
for D and I-WP. Since P and G are related to D by a Bonnet
transformation, their distributions can be obtained simply by
appropriately rescaling. For example, for the transformation $D \to G$,
the $K_i$ have to be rescaled with $a_G^2 / a_D^2$ and the $f_i$ with
$A_G^* / A_D^*$, as given in Tab.~\ref{TableNumbers}.  In
Fig.~\ref{FigureDistributions1}a, we plot the four different
distributions obtained. P, D and G feature one peak around a
(relatively large) K-value, that means they are quite uniformly
curved, although they have flat regions, too (which correspond to the
poles of $R(z)$). I-WP is bimodal, that is it has prominently both
flat and strongly curved regions. The strong curvature of D is somehow
artificial due to our choice of unit cell.  If one corrects for this
by a factor $1/4$ for the K-values, it becomes clear that I-WP has a
much broader distribution than P, D and G.  This can also be inferred
from Tab.~\ref{TableMoments}, where we list the first moments $S_N$ of
the distributions for D and I-WP. The ones for D are identical to the
ones given in Ref.~\cite{a:ande88} after appropriate rescaling $S_N$
with $8^{-(2 N+1)/3}$ due to the different choice in unit cell (the
numerical deviation is less than 0.02 \% even for $N = 8$). The
moments for P and G can be calculated from the ones for D due to the
Bonnet-transformation between them; e.g.\ $S_N^G = S_N^D (N_G / N_D)
(a_G / a_D)^{2 (N-1)}$ with the values for $N$ and $a$ given in
Tab.~\ref{TableNumbers}.  In particular, $A^*_G = A^*_D (N_G a_D^2) /
(N_D a_G^2)$, $\chi_G = \chi_D (N_G / N_D)$ and $\Gamma_G = \Gamma_D
(N_G a_D^3)/(N_D a_G^3) $ --- as can be verified from
Tab.~\ref{TableNumbers}. It follows immediately that the quantity
$\Xi(K^*) = K^* A^* / 8 \pi \chi$ defined for
\eq{FreeEnergyDensityBicont} is invariant under a Bonnet
transformation.

In order to quantify the variance of the different distributions, we
calculate standard deviation, $\Delta = \langle ( K - \langle K
\rangle )^2 \rangle / \langle K \rangle^2 = S_0 S_2 / S_1^2 - 1$ (here
$\langle \dots \rangle$ means area average). This quantity is
independent of scaling, choice of unit cell and Bonnet-transformation,
and given in the last row of Tab.~\ref{TableMoments}. As expected, its
value for I-WP is higher than for P/D/G.

\section{Curvature properties from Ginzburg-Landau}
\label{sec:GinzburgLandau}

In order to investigate the curvature properties of S, C(P) and F-RD,
we use the representations as isosurfaces to fields $\Phi(\textbf{r})$ which
we obtained recently as local minima of a simple Ginzburg-Landau
theory for ternary amphiphilic systems \cite{uss:schw99}. The same
model can also be employed for binary systems if
the scalar order parameter field $\Phi(\textbf{r})$ is interpreted to be the
local concentration difference between water in the first labyrinth
and water in the second labyrinth. Then the mid-surface of the lipid
bilayer is identified with the surface $\Phi(\textbf{r})=0$. The free-energy
functional is given by
\begin{equation} \label{Phi6}
{\cal F}[\Phi] = \int\ d\textbf{r}\ \left\{ (\Delta \Phi)^{2}
+ g(\Phi) (\nabla \Phi)^{2} + f(\Phi) \right\}\ .
\end{equation}
A reasonable choice for $f$ and $g$ has been found to be 
\begin{equation} \label{GL_density}
f(\Phi) = (\Phi + 1)^{2} (\Phi - 1)^{2} (\Phi^{2} + f_0)\ ,\ \qquad
g(\Phi) = g_0 + g_2 \Phi^{2}\ .
\end{equation}
For given model parameters $(g_0, g_2, f_0)$, the free-energy functional is
minimized in Fourier space by implementing the correct space group
symmetry and minimizing for Fourier amplitudes and lattice constant
with conjugate gradients. For balanced TPMS, an additional black and
white symmetry has to be implemented.  The resulting fields
$\Phi(\textbf{r})$ correspond to the various local minima. We showed
in Ref.~\cite{uss:schw99} that by tuning $(g_0, g_2, f_0)$
appropriately, the $\Phi(\textbf{r}) = 0$-isosurfaces can be made to
become very close to minimal surfaces, and generated representations
for nine different TPMS: G, D, P, S, C(P), C(D), C(Y), I-WP and F-RD.

K-distribution can be obtained from these representations as follows.
By triangulating the isosurfaces with the marching cube algorithm, we
numerically obtain a metric on the surface. From the Fourier
representations of the fields $\Phi(\textbf{r})$, it is possible to
calculate exactly the K-values on the vertices of the triangulation.
Using a histogram technique like in the Weierstrass case, it is thus
possible to obtain $\{K_i, f_i\}$ even for structures like S, C(P) and
F-RD, which might be of physical relevance but for which no exact
representations are known. The distributions obtained in this way for
$g_0 = -3$, $g_2 = 7.01$ and $f_0 = 0$ are plotted in
Fig.~\ref{FigureDistributions1}b.  All three of them are roughly
trimodal and feature much stronger curved parts than the ones
discussed above. This reflects their complicated shape, which makes
them more difficult to access numerically than the ones which have been
generated by Weierstrass representations; it also explains the
difficulty to obtain smooth data for the K-distributions.  In
Tab.~\ref{TableMoments}, we give their first moments $S_N$ and the
measure $\Delta$ of their variance. All of them have broader
distributions than I-WP, in the sequence S - F-RD - C(P). Therefore we
can conclude already from the distributions that the IBCPs should become
less favorable in the sequence P/D/G - I-WP - S - F-RD - C(P).

\section{Phase behavior}
\label{sec:PhaseDiagrams}

\begin{figure}
\begin{center}
\leavevmode 
\psfig{file=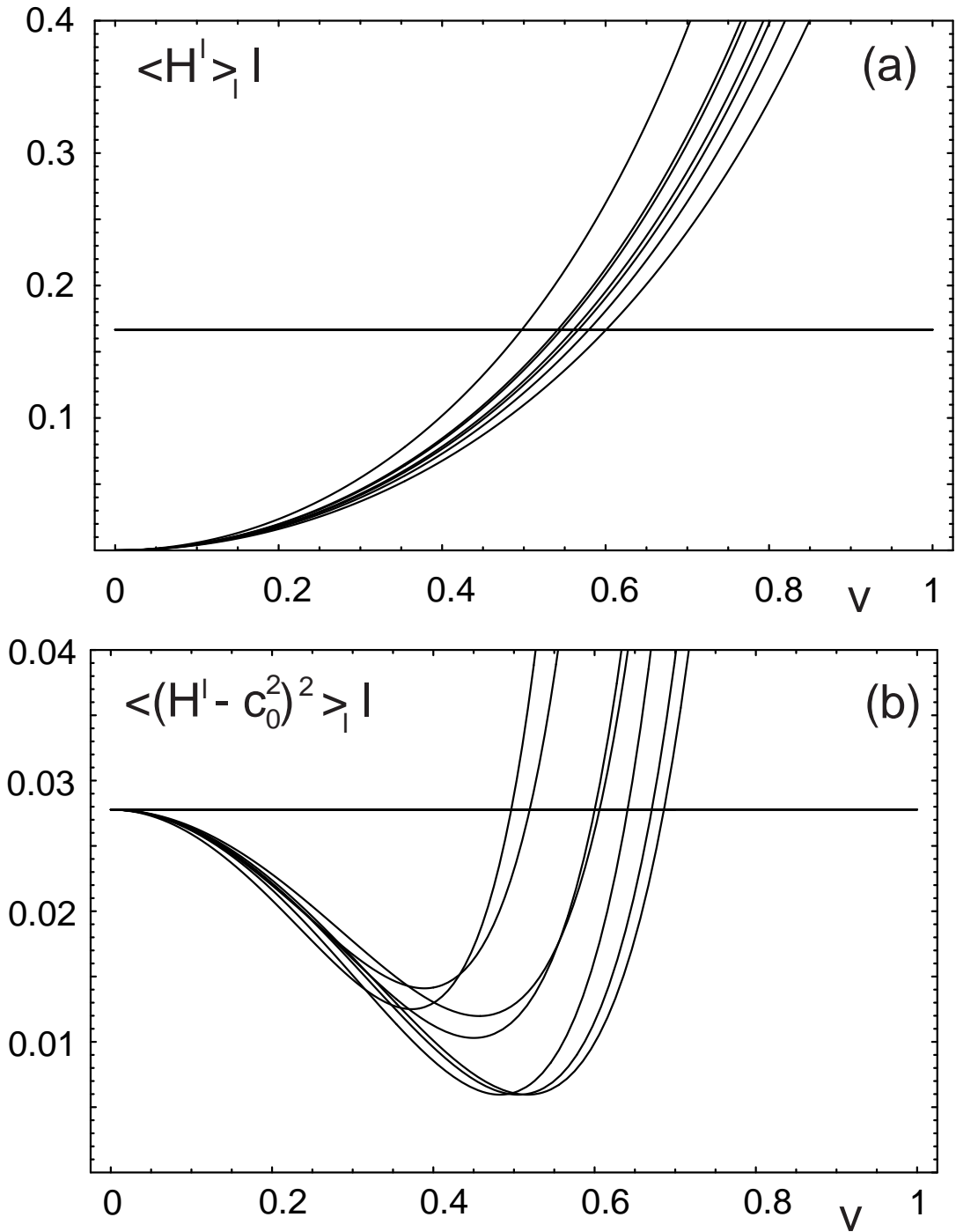,width=0.8\textwidth}
\end{center}
\caption{(a) $\langle H^l \rangle_l l = (v/\Gamma)^2/(4 - (v/\Gamma)^2)$ as a
  function of hydrocarbon volume fraction $v$. The straight line is $c_0 l
  = 1/6$.  From right to left, the IBCPs appear in the sequence 
  S, G, D, I-WP, P, C(P), and F-RD. (b) $\langle (H^l - c_0)^2 \rangle_l l^2$ 
  as a function of hydrocarbon volume fraction $v$ with $c_0 l = 1/6$.  
  The IBCPs can be identified from their sequence at the top
  of the figure, where G, D, P, S, I-WP, C(P) and F-RD appear from
  right to left. The straight line is
  $(c_0 l)^2 = (1/6)^2$.  Since P, D and G are related by a Bonnet
  transformation, their curves differ from each other only by a simple
  rescaling of $v$ with $\Gamma$.  Therefore they reach the same
  minimal values, but at different $v$ according to their $\Gamma$-values.}
\label{FigureGeometry1}   
\end{figure}

Given the distribution of Gaussian curvature as histogram
$\{K_i, f_i\}$, it is straightforward to evaluate the first term
of the curvature energy, \eq{FreeEnergyDensityBicont2}, of the bicontinuous 
phases. Since it is an area average over a
complicated function of $K^*$, we have to replace $\int dA^*$ and
$K^*$ by $\sum f_i$ and $K_i$, respectively.  In
Fig.~\ref{FigureGeometry1}b, we plot the quantity $\langle (H^l -
c_0)^2 \rangle_l l^2$ as a function of hydrocarbon volume fraction $v$
for $c_0 = 1/6$.  In contrast to $\langle H^l \rangle_l l$, compare
\eq{eq:GeometricalProperty}, no simple formula
exists, and it has to be calculated numerically as
\begin{equation} \label{eq:GeometricalProperty3}
\langle (H^l - c_0)^2 \rangle_l l^2 =
\frac{\sum_i f_i \left( 1 - \Xi_i \left(\frac{v}{\Gamma} \right)^2 \right) 
\left( \frac{\Xi_i \left(\frac{v}{\Gamma} \right)^2}
{1 - \Xi_i \left(\frac{v}{\Gamma} \right)^2} + c_0 \right)^2} 
{\sum_i f_i (1 - \Xi_i \left(\frac{v}{\Gamma} \right)^2)} \ .
\end{equation}
Since $\Xi_i$ is invariant under a
Bonnet transformation, $\langle (H^l - c_0)^2 \rangle_l l^2$ is the
same for P, D and G, except for a rescaling of $v$ with $\Gamma$. In
particular, the curves reach the same minimal values and the minima are
shifted to the right with increasing topology index $\Gamma$. Thus the
structures P, D and G perform equally well, but at different values of
$v$. From Fig.~\ref{FigureGeometry1}a, we see that all other structures
cannot reach this low value of frustration since they have much
broader distributions than P, D and G (compare
Tab.~\ref{TableMoments}). Note that the scaling with $v / \Gamma$ is
valid in \eq{eq:GeometricalProperty} for all structures and in
\eq{eq:GeometricalProperty3} only for those related by a Bonnet
transformation.

\begin{figure}
\begin{center}
\leavevmode 
\psfig{file=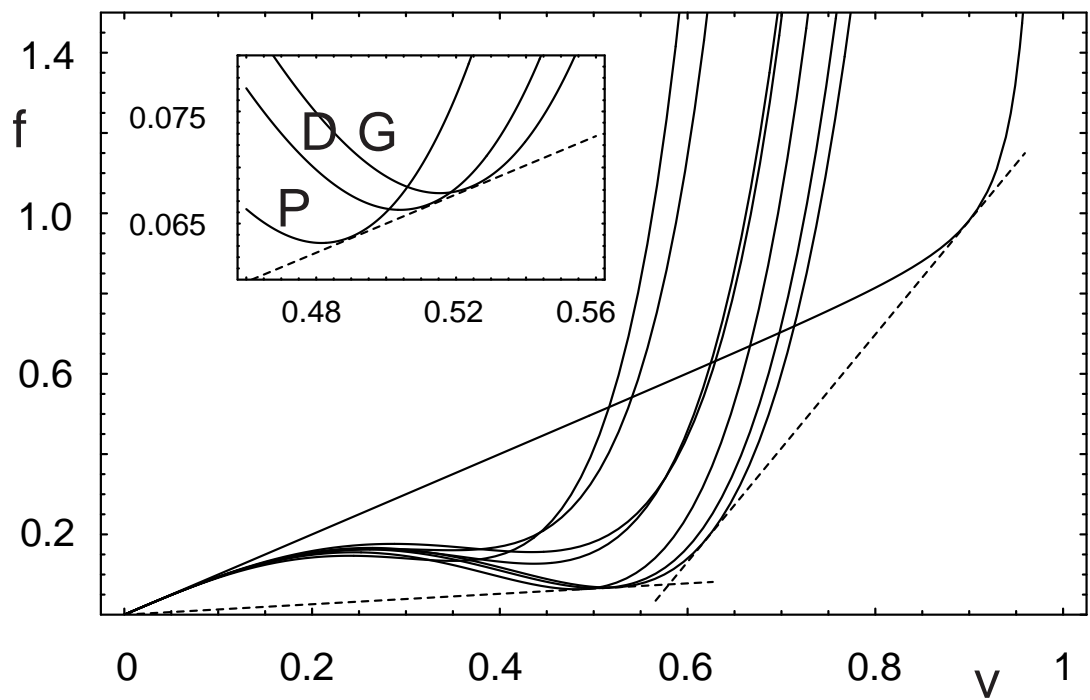,width=0.8\textwidth}
\end{center}
\caption{Free energy densities as a function of hydrocarbon volume fraction 
  $v$ for $c_0 l = 1/6$, $r = 0$ and $\kappa/k_B T = 10$. The solid
  line on the right is $f_L$, the other solid lines are the different
  $f_b$. The IBCPs can be identified from their sequence at the top of
  the figure, where G, D, P, S, I-WP, C(P) and F-RD appear from right
  to left. With decreasing $v$, the phases $L_{\alpha}$, G, D and P
  are stable. The lower dashed line is the Maxwell construction with
  an excess water phase (emulsification failure). G, D and P lie on a
  triple line (compare inset) since they are related by a Bonnet
  transformation. The right dashed line is the Maxwell construction
  between G and $L_{\alpha}$. Taken from
  Ref.~\protect\cite{letter}.}
\label{FigureFreeEnergies}   
\end{figure}

Except for a factor of $v / c_0^2$, the free-energy density 
(\ref{FreeEnergyDensityBicont2}) for $r=0$ is identical with $\langle
(H^l - c_0)^2 \rangle_l l^2$ (see \eq{eq:GeometricalProperty3}).  In
Fig.~\ref{FigureFreeEnergies}, we show the free-energy density of all
considered phases for $c_0 = 1/6$, $r=0$ and $\kappa / k_B T = 10$.
>From the graph, we infer the phase sequence L - G - D - P with
increasing water content; the other phases are not stable since they
cannot reach that small an amount of bending frustration as can P, D
and G. As discussed above, \eq{eq:MeanCurvatureAtc0} implies that the 
sequence G - D - P is determined by the topology index.

Since the free-energy density for P rises again for small $v$, it
would be more favorable for it to coexist with excess water.  This
mechanism is known as \emph{emulsification failure} in surfactant
systems \cite{a:safr99} and allows to extend the sequence to include
a P+W coexistence. The emulsification failure can be interpreted as a
Maxwell construction of P with the pure water phase with $f = 0$ at $v =
0$. It follows from \eq{FreeEnergyDensityBicont2} and the scaling of $\langle
(H^l - c_0)^2 \rangle_l l^2$ with $v / \Gamma$ that the
free-energy density $f_b$ for P, D and G can be written in the scaling form 
\begin{equation} \label{eq:scaling}
f_b = v\, \Omega(v/\Gamma)
\end{equation}
with a universal function $\Omega(x)$. For small $r$, this
function has $\Omega(0) = 1$, a minimum at $x \simeq 4 c_0 / (1+c_0)$, and
diverges at $x \simeq 4$. A Maxwell construction shows that these
structures always lie on a triple line, irrespective of the values for
$c_0$, $r$ and $\kappa$ (compare inset of
Fig.~\ref{FigureFreeEnergies}).  This means that if we calculated the
whole phase diagram with the Maxwell construction, D and P would be
stable only along a line.  In the following we identify phase
transitions between G, D and P with the intersections of the
free-energy curves, since under experimental conditions, small
additional contributions to the free energy are certain to destroy the
delicate free-energy balance resulting from the Bonnet transformation,
 and then will lead to extended one-phase regions for D and P 
(or remove one or both of them from the overall phase diagram).

\begin{figure}
\begin{center}
\leavevmode
\psfig{file=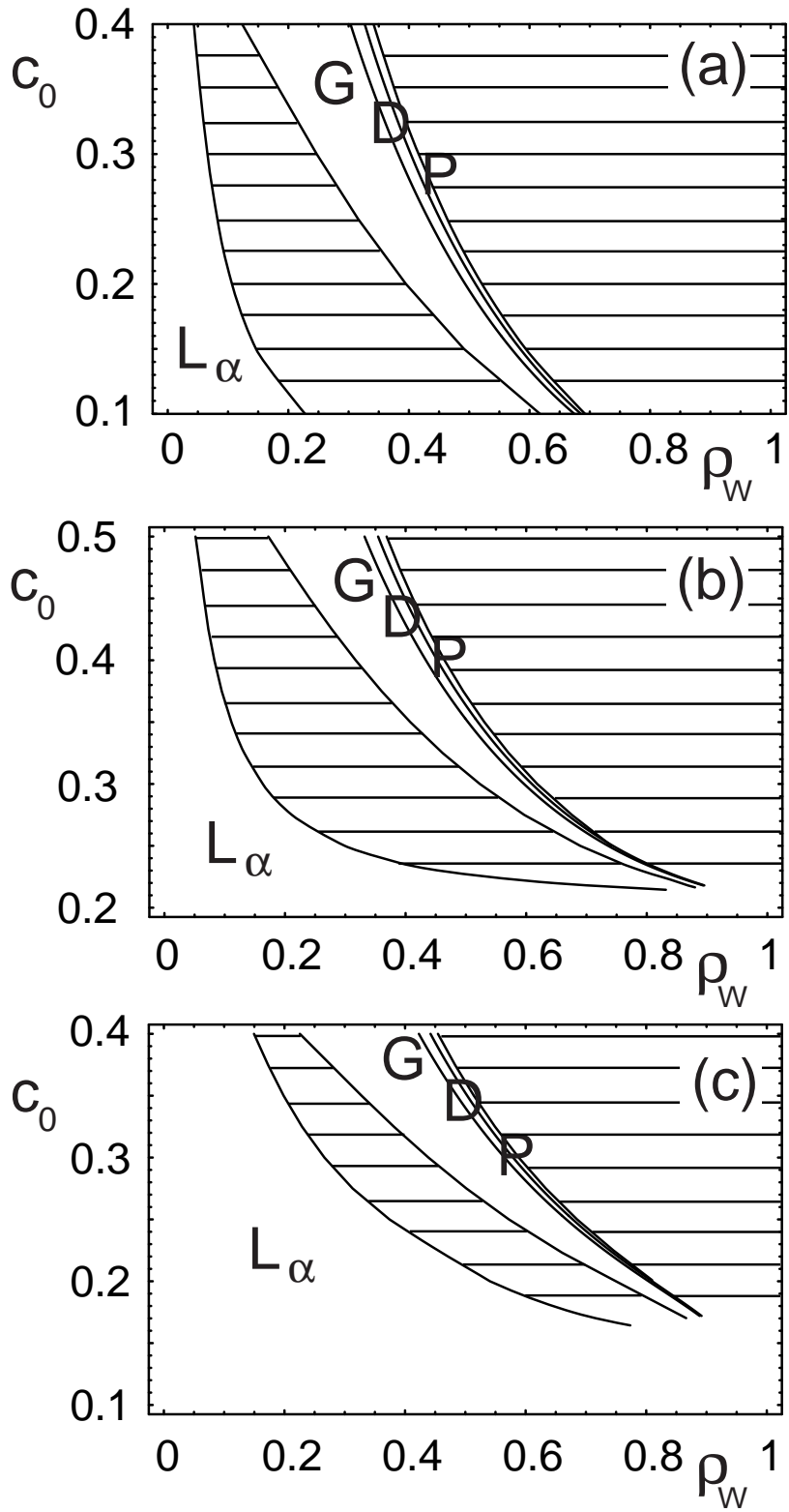,width=0.6\textwidth} 
\end{center}
\caption{Phase diagrams as a function of water volume fraction 
  $\rho_W=1-v$ and spontaneous curvature $c_0$ for (a) $r = 0.1$ and
  $\kappa/k_B T = 10$, (b) $r = 0.5$ and $\kappa/k_B T = 10$ and (c)
  $r = 0.5$ and $\kappa/k_B T = 2$.  Two-phase coexistences are
  indicated by hatched regions. (a) has been presented before in
  Ref.~\protect\cite{letter}.}
\label{FigurePhaseDiagram}
\end{figure}

In \fig{FigurePhaseDiagram} we use this reasoning to construct phase
diagrams as a function of $v$ and $c_0$ for different values of $r$
and $\kappa / k_BT$. With increasing water concentration $\rho_W=1-v$
we always obtain the sequence L - G - D - P - W+P.  The most stable
IBCP is the gyroid structure G close to the lamellar phase at high
values of $v$ since it has both a narrow distribution of Gaussian
curvature and the highest topology index. At lower concentrations of
$v$, one finds narrow regions of stability for D and P since they have
narrow distributions, too, but increasingly lower values of $\Gamma$.
Increasing $c_0$ shifts the loci of least frustration to higher $v$ as
evident from Fig.~\ref{FigureGeometry1}a. Decreasing $\bar \kappa$
(i.e. increasing $r$) disfavors the IBCPs and increases the
stability region for the lamellar phase, as can be seen by comparing
\fig{FigurePhaseDiagram}a and \ref{FigurePhaseDiagram}b, where $r$ is 
increased from $0.1$ to $0.5$, respectively.  However, for 
$c_0 \gtrsim r/2$ there always will be a region
of stability for the IBCPs, even for $r = 1$. This stands in marked
contrast to similar modelling for ternary systems where bicontinuous
cubic phases are expected to disappear for $r \gtrsim 0.55$
\cite{uss:schw00a}. The explanation lies in the fact that changing the
spontaneous curvature $c_0$ of the monolayers amounts to changing the
effective saddle-splay modulus of the bilayer, compare 
Eq.~(\ref{eq:EffectiveBilayerBending}).
Thus, increasing $c_0$ can counteract the effect of making $\bar
\kappa$ more negative. The bicontinuous phases are also favored by
lowering $\kappa$ without changing $r$, since this increases the
fluctuation effects which favor IBCPs (positive renormalization of
$\bar \kappa$) and disfavor the lamellar phase (increase of steric
repulsion). This can be seen by comparing \fig{FigurePhaseDiagram}b
and \ref{FigurePhaseDiagram}c, where $\kappa / k_B T$ is decreased from 
$10$ to $2$, respectively, while $r$ is kept constant at $0.5$.

\section{Breaking the Bonnet Symmetry}
\label{sec:symmetry_breaking}

It is clear from our discussion in Sec.~\ref{sec:Model} that all
contributions to the free energy, which can be written in the scaling
form $v\, \Omega(2 l \langle K \rangle^{1/2}) = v\, \Omega(v/\Gamma)$,
will preserve the triple-line coexistence of G, D and P. Obviously, an
interaction which introduces an additional length scale, in addition
to $l$ and $\langle K \rangle^{-1/2}$, is needed to break the Bonnet
symmetry and to lift the triple-line coexistence. Here we want to
discuss van der Waals interactions and chain stretching. 

The non-retarded van der Waals interaction at distance $d=|{\bf r}_1 -
{\bf r}_2|$ has the general form
\begin{equation} \label{eq:vdW_interaction}
w(d) = - \frac{A_H}{\pi^2} \rho_1 \rho_2\ d^{-6} 
\end{equation}
where $A_H$ is the Hamacker constant, and $\rho_1$ and $\rho_2$ are
the number densities of atoms in the two interacting bodies.  The
total interaction energy between two colloidal particles of volumes
$V_1$ and $V_2$ is then given by
\begin{equation} \label{eq:vdW_energy}
W = \int_{V_1} d^3 r_1 \int_{V_2} d^3 r_2\ w(|{\bf r}_1-{\bf r}_2|)\ .
\end{equation}
Unfortunately, the energy density of the van der Waals interaction
cannot be calculated easily in a cubic bicontinuous phase. However, it
is clear from Eq.~(\ref{eq:vdW_energy}) that it is the embedding of
the TPMS in three-dimensional space, which is important for the van
der Waals energy. Distances between different parts of the bilayer are
not determined uniquely by the $K$-distribution, which describes the
internal geometry of a surface. This is evidenced by the
self-intersections of the parallel surfaces for $v \gtrsim 0.8$, which
do {\em not} affect any of the considerations of the curvature energy
described in the previous sections, but would give a divergent
contribution to the van der Waals energy. Thus, we conclude that the
van der Waals interaction must break the Bonnet symmetry. We
illustrate this point in Appendix~\ref{app:FCC}, where we approximate
the TPMS by a FCC array of spherical shells.

In order to estimate the frustration energy due to chain stretching,
it has been suggested in Ref.~\cite{a:ande88} to calculate the
variance of the distance between constant-mean-curvature companions of
a given TPMS.  For the D-surface, the variance $\Delta_L = \langle
(L-l)^2 \rangle$ has been shown \cite{a:ande88} to be almost independent 
of $v$, with
\begin{equation} \label{DeltaL}
\Delta_L = \alpha_D v^2 a^2 = 4 \alpha_D  {A^*}^2 l^2 
\end{equation}
and $\alpha_D = 0.00035$, where Eq.~(\ref{LatticeConstant}) has been
employed in the second equality. Similar results can be expected for G
and P, with prefactors $\alpha_G$ and $\alpha_P$, respectively.  We
introduce a function $\alpha(A^*)$, which is defined such that
$\alpha(A^*_P)=\alpha_P$, $\alpha(A^*_D)=\alpha_D$ and
$\alpha(A^*_G)=\alpha_G$ for the three values of $A^*$ for the P, D
and G phases.  Then, the stretching-energy density has the scaling
form
\begin{equation} \label{eq:stretch_scaling}
f_s = \frac{l^3}{2\kappa c_0^2} \ \frac{E_s}{V} = 
         v \, \frac{k_s(l)l^4}{2\kappa c_0^2} \, \Delta_L l^{-2} = 
    2 \alpha(A^*)  {A^*}^2 \, \frac{k_s(l)l^4}{\kappa c_0^2} \, v 
\end{equation}
where $E_s$ is the stretching energy from \eq{eq:stretching} and
\eq{DeltaL} has been used for $\Delta_L$. Note that this energy
density has been made dimensionless with the same factor as
the curvature energy in Sec.~\ref{sec:Model}, i.e. it is normalized
to the free-energy density of the lamellar phase 
[compare Eq.~(\ref{FreeEnergyDensityLamellar})]. 

These results make the scaling argument of Sec.~\ref{sec:Model} more
precise.  Equation~(\ref{DeltaL}) shows very clearly that the lattice
constant $a$ appears as new length scale, and that the Bonnet symmetry
is broken in general by the stretching contributions.  It is
interesting to consider the possible functional dependencies of
$\alpha(A^*)$ in more detail. First, if $\alpha(A^*){A^*}^2$ is a
monotonically increasing (decreasing) function of $A^*$, then P is
stabilized (destabilized) by chain stretching due to its small value
of $A^*$, while D is disfavored (favored). In the case of increasing
$\alpha(A^*){A^*}^2$, D would not appear in the phase diagram, while
in the case of decreasing $\alpha(A^*){A^*}^2$, P cannot be present.
Second, if $\alpha(A^*){A^*}^2=$ const, the Bonnet symmetry is
restored.  Finally, $\alpha(A^*){A^*}^2$ can of course be a
non-monotonic function of $A^*$, which stabilizes the phase with the
smallest value of $\alpha(A^*){A^*}^2$.
 
In any case, the difference between the three phases can be expected
to be small, since the prefactor $\alpha_D$ (and presumably also
$\alpha_G$ and $\alpha_P$) is very small; nevertheless, the stretching
energy might account for small differences, like the ones needed to
produce the phase diagram of 2:1 LA/DLPC and water, which shows no
triple-line symmetry, but something close to it.  A more precise
statement requires the calculation of $\Delta_L$ for G and P in the
framework of the CMC-model. Other estimates of the stretching energy
have been discussed in Ref.~\cite{a:temp98a}; they also indicate a
breaking of the Bonnet symmetry. 

\section{Concluding Remarks}
\label{sec:Discussion}

In this work, we have investigated inverse bicontinuous cubic phases
(IBCPs) in lipid-water mixtures. We systematically included all IBCPs
which might be expected to be of physical relevance, and found that
only the ones with a narrow distribution of Gaussian curvature over
the minimal mid-surface are stable, i.e.\ P, D and G.  While this
result has been anticipated in Ref.~\cite{a:helf90}, we were able to
prove it quantitatively for the first time by calculating the
distributions of Gaussian curvature over the minimal mid-surfaces of
all relevant structures; from these we calculated the variance 
$\Delta$ of the $K$-distributions given in \tab{TableMoments} which is
the same for P, D and G --- due to Bonnet transformations --- and
higher for all other structures.

The Bonnet transformation also implies that the free-energy densities
for P, D and G have the scaling form $f_b = v\, \Omega(v / \Gamma)$
with hydrocarbon volume fraction $v$ and topology index $\Gamma$. 
The topology index is a dimensionless inverse Euler characteristic
which is independent of scaling and choice of unit cell; the higher
its value, the less porous the structure. Since it is highest for the
gyroid structure G, it has the largest region of stability. D and P
have progressively lower values for $\Gamma$, and have therefore
smaller regions of stability at lower values for $v$.

We also used the concept of the emulsification failure (which is well
established in surfactant science) to explain the stability of IBCPs
in excess water. A Maxwell construction with the excess phase shows
that P, D, G and excess water lie on a triple line for all values of
$c_0$ and $r$ as a consequence of the Bonnet transformation.  This
'Bonnet symmetry' should not be interpreted as a degeneracy between G,
D and P, since their different topologies ensures that their regions of
stability are arranged according to their respective values of the
topology index. It should be noted that the free energy per {\em unit 
area} (rather than per unit volume) has the scaling form 
$\Omega(v/\Gamma)$; this might be interpreted as some kind of 
degeneracy (compare Refs.~\cite{a:helf90,a:temp98a}), since now all 
three structures attain the same minimal values.

\begin{figure}
\begin{center}
\leavevmode
\psfig{file=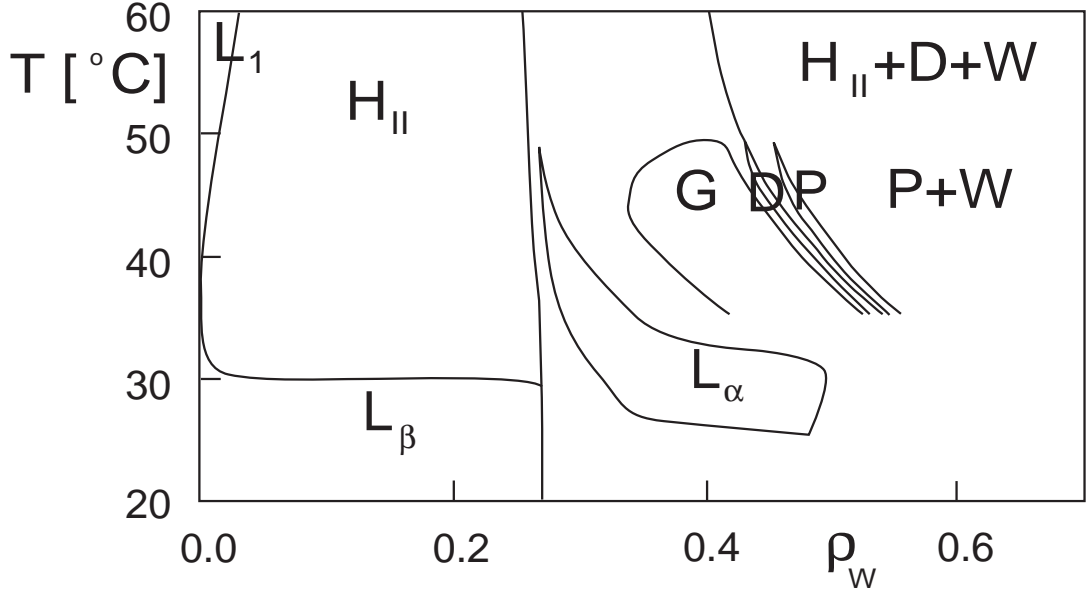,width=0.8\textwidth}
\end{center}
\caption{Experimental phase diagram for 2:1 lauric acid/dilauroyl 
  phosphatidylcholine and water as a function of water volume fraction
  $\rho_W$ and temperature $T$ (redrawn schematically from 
  Ref.~\protect\cite{a:temp98a}). The temperature range from 
  $T = 35^0 C$ to $T = 50^0 C$ corresponds roughly to the range $0.18$ 
  to $0.26$ in spontaneous curvature $c_0$ in the predicted phase
  diagram, compare \protect\fig{FigurePhaseDiagram}a.}
\label{Seddon}
\end{figure}

Our results agree nicely with experimental observations. At fixed
temperature, it is always the sequence G - D - P which is observed
with increasing water content, with G having the largest region of
stability and only D or P coexisting with excess water
\cite{a:temp98}.  For example, for monoolein and water at $40^0 C$,
with increasing water concentration the sequence L - G - D has been
found, where G has the largest region of stability and D coexist with
excess water \cite{a:hyde84}. Our phase diagram for $r = 0.1$ and
$\kappa / k_B T = 10$ (\fig{FigurePhaseDiagram}a)
agrees particularly well with the one for 2:1
LA/DLPC and water \cite{a:temp98a} (compare \fig{Seddon}). We do not
treat the inverse hexagonal phase, which experimentally occurs at
large $v$ and large $c_0$, but we showed above that IBCPs will occur
at large $c_0$ and {\em small} $v$, and argued that at large $v$
stretching energy should become comparable to bending energy. This, in
fact, favors the inverse hexagonal phase, whose geometry requires
chain stretching.

Finally, we can discuss the role of temperature for 2:1 LA/DLPC and
water by using the ansatz $c_0 = b (T - T_b)$, with a balanced
temperature $T_b$ and a phenomenological constant $b$. Using $l =
11.4\ \AA$, we find $b \approx 0.0011 / \AA K$ and $T_b \approx 0^o
C$. Since the chain melting temperature is known to be $T_m = 30^o C$,
we conclude that the balanced temperature $T_b$ is well below the main
transition. This seems to be the generic case for lipid systems; it
might be related, e.g., to the anomalous swelling of PC-bilayers when
the main transition is approached from above \cite{a:rich99}.  It is
interesting to note that the constant $b$ obtained here for 2:1
LA/DLPC and water is very close to the corresponding value $b=0.0012 /
\AA K$ for monolayer of $C_{12} E_5$ in ternary microemulsions
\cite{a:stre94}. 

In this paper, we only considered inverse bicontinuous phases with
cubic symmetry. A similar analysis for non-cubic phases should be
straight-forward, since here Weierstrass representations are known for
the main structures, especially for the rhombohedral and tetragonal
variants of G, D and P \cite{a:fodg99}.  However, we expect that these
phases will not be more favorable than G, D and P for the same reason
as for the other cubic phases. In order to make further progress in
understanding their phase behavior, it seems more important to
investigate the different mechanisms which might break the Bonnet
symmetry between G, D and P.  In our view, two calculations are needed
now: (i) of the van der Waals energy of G, D and P and (ii) of the
variance of the distance distribution of a constant-mean-curvature
model of G and P. 

\begin{appendix} 
\section{Estimate of van der Waals Energy Density}
\label{app:FCC}

For a rough estimate of the energy density due to the van der Waals 
interaction in cubic bicontinuous phase, we use a FCC array of 
spherical shells of radius $R$, thickness $2l$ and lattice constant 
$a$.  Here, the lattice constant $a$ corresponds to the
conventional unit cell and is taken to be the same as for the
corresponding bicontinuous phase, and $R$ is determined by the
requirement of equal area $A$ per unit cell, so that $A^* = 16 \pi
(R/a)^2$. Thus we have $a/l = 2 A^* / v$ again. The distance of
closest approach between two neighboring spheres is $D = \sqrt{2} a /
2 - 2 R$; in order to avoid contact between them, $D > 0$ is
necessary, which implies $A^* < 2 \pi$; this inequality is satisfied
for all TPMS studied here, compare Tab.~\ref{TableIndex}.  For the
crystal, we consider the interaction between nearest-neighbor spheres
only. The van der Waals energy between two hollow spheres is described
by the Girifalco potential, which is often used to model the van der
Waals interaction between two buckyballs $C_{60}$ \cite{n:giri92}.
For our purpose it is sufficient to use the Derjaguin approximation for 
the Girifalco potential, which describes the leading contributions for 
close approach correctly, and shows that the van der Waals attraction for 
two hollow spheres scales as $W \sim A_H R l^2 / D^3$. We thus find for 
the van der Waals energy density
\begin{equation}
f_{vdW} = \frac{l^3}{2\kappa c_0^2}\ \frac{W}{V} 
   \sim \frac{A_H}{\kappa c_0^2} \left( \frac{l}{a} \right)^5 
   \left( \frac{R}{a} \right)
   \left( \frac{1}{\sqrt{2}} - \left( \frac{R}{a} \right) \right)^{-3}\ .
\end{equation}
This energy
density has again been made dimensionless with the same factor as
the curvature energy in Sec.~\ref{sec:Model}. 
Its scaling shows that {\em two} ratios of
length scales, $R/a\sim (A^*)^{1/2}$ and $a/l\sim A^*/v$, determine
the van der Waals energy.

It is intersting to note that the curvature-energy density in this model 
of spheres scales as
\begin{equation} \label{BendingSphere}
f_b = v \left[1 - \frac{1}{c_0 (R/l)}\right]^2\ . 
\end{equation}
Since $R/l \sim (A^*)^{3/2} / v \sim \Gamma/v$ (for fixed $\chi$),
\eq{BendingSphere} is consistent with the scaling function
(\ref{eq:scaling}). This indicates that our identification of the 
sphere radius and lattice constant gives a consistent approximation
for both van der Waals and bending energies. 

\end{appendix}

\begin{table}
\begin{center}
\begin{tabular}{|l|l|l|l|l|l|l|l|} \hline
         & S & G & D & I-WP & P & C(P) & F-RD \\ \hline
$\chi$   & -40 & -8 & -16 & -12 & -4 & -16 & -40 \\ \hline  
$A^*$    & $5.41457$ & $3.09140$& $3.83755$ & $3.46367$ & $2.34516$ & $3.74820$ & $4.77522$ \\ \hline
$\Gamma$ & $0.794735$ & $0.76665$ & $0.74978$ & $0.74238$ & $0.71637$ & $0.655994$ & $0.654174$ \\ \hline
\end{tabular}
\end{center}
\caption{Euler characteristic $\chi$, scaled surface area $A^*$ 
(both in the conventional unit cell) and topology index $\Gamma = 
({A^*}^3/2 \pi |\chi|)^{1/2}$ for all TPMS considered, as obtained from 
local minima of the free-energy functional
of a simple Ginzburg-Landau theory.  Exact values
are known for G, D, I-WP, P and C(P) and given in \tab{TableNumbers}.
The structures are ordered according to decreasing $\Gamma$; the
value for S is probably too large, but difficult to improve numerically.}
\label{TableIndex}
\end{table}  

\begin{table}
\begin{center}
\begin{tabular}{|l|l|l|l|l|l|} \hline
     & $\chi$ & $A^*$ & $\Gamma$ & $N$ & $a$  \\ \hline
G    & $-8$ & $3 (1 + k_2^2) / 2 k_2 = 3.091444$ & $0.766668$ & 24  
     & $8 k_1 / 3 \sqrt{1+k_2^2} = 2.656243$  \\ \hline
D    & $-16$ & $3 / k_2 = 3.837785$ & $0.749844$ & 48 & $8 k_1 / 3 = 3.37150$ \\ \hline
I-WP & $-12$ & $2 \sqrt{3} = 3.464102$ & $0.742515$ & 48 
     & $3 \sqrt{3} k_4^3 / 4 \pi= 7.949874$ \\ \hline
P    & $-4$ & $3 k_2 = 2.345103$ & $0.716346$ & 12 & $4 k_1 / 3 k_2 = 2.156516$ \\ \hline
C(P) & $-16$ & $3/k_3 = 3.510478$ & $0.655993$ & - & - \\ \hline
\end{tabular}
\end{center}
\caption{Scaled surface area $A^*$, 
Euler characteristic $\chi$ in the conventional unit cell,
and topology index $\Gamma = ({A^*}^3/2 \pi |\chi|)^{1/2}$ for those TPMS,
for which exact results are available.  For surfaces with known Weierstrass
representations, we also give $N$, the number of fundamental domains 
needed to build up the surface in the conventional unit cell,
and $a$, the resulting lattice constant. Here $k_1 = F(\sqrt{3}/2,\sqrt{8}/3)$ 
where $F$ is the incomplete elliptic integral of the first kind;
$k_2 = K(1/2) / K(\sqrt{3}/2)$ where $K(k) = F(1,k)$ is the complete 
elliptic integral of the first kind; $k_3 = K(1/\sqrt{3}) / K(\sqrt{2/3})$; 
and $k_4 = \Gamma(1/3)$.   
The surface areas and Euler characteristics of P, D and G are related to
each other due to their Bonnet-transformations, as described in the text. 
The Bonnet angle $\theta = 38.015^o$ 
for G follows from  $\tan \theta = k_2$. Note that often a smaller unit
cell is chosen for D; then one has $\chi = -2$ and $A^* = 1.9188925$.}
\label{TableNumbers}
\end{table}  

\begin{table}
\begin{center}
\begin{tabular}{|l|l|l|l|l|l|} \hline
& D & I-WP & S & F-RD & C(P) \\ \hline \hline
$S_0$ 
& $3.83763$ 
& $3.46250$ 
& $5.41457$
& $4.77522$
& $3.74820$ \\ \hline
$S_1$ 
& $- 100.531$ 
& $- 75.4075$ 
& $- 251.403$
& $- 251.316$
& $- 100.992$ \\ \hline
$S_2$  
& $3209.46$ 
& $2434.91$ 
& $18514.0$
& $21821.3$
& $5012.37$ \\ \hline
$S_3$ 
& $- 110842$ 
& $- 86861.5$  
& $- 1.72204 \times 10^6$
& $- 2.29671 \times 10^6$
& $- 286143$ \\ \hline
$S_4$  
& $4.00533 \times 10^6$ 
& $3.24503 \times 10^6$
& $1.82465 \times 10^8$
& $2.66794 \times 10^8$
& $1.81085 \times 10^7$ \\ \hline
$S_5$  
& $- 1.49309 \times 10^8$ 
& $- 1.24566 \times 10^8$  
& $- 2.08738 \times 10^{10}$
& $- 3.27706 \times 10^{10}$
& $- 1.21507 \times 10^9$ \\ \hline
$S_6$  
& $5.69880 \times 10^9$ 
& $4.86942 \times 10^9$  
& $2.50549 \times 10^{12}$
& $4.16686 \times 10^{12}$
& $8.45482 \times 10^10$ \\ \hline
$S_7$  
& $- 2.21675 \times 10^{11}$ 
& $- 1.92889 \times 10^{11}$
& $- 3.10619 \times 10^{14}$
& $- 5.42345 \times 10^{14}$
& $- 6.02391 \times 10^{10}$ \\ \hline
$S_8$  
& $8.75987 \times 10^{12}$ 
& $7.719305 \times 10^{12}$
& $3.941396 \times 10^{16}$
& $7.179904 \times 10^{16}$
& $4.363058 \times 10^{14}$ \\ \hline \hline
$\Delta$ 
& $0.218702$
& $0.482666$
& $0.586079$
& $0.649801$
& $0.842022$ \\ \hline 
\end{tabular}
\end{center}
\caption{First moments $S_N$ and variance $\Delta = S_0 S_2 / S_1^2 - 1$ 
of the distributions of Gaussian curvature.
In particular, $S_0 = A^*$ and $S_1 = 2 \pi \chi$ with $A^*$ and $\chi$ 
given in Tab.~\ref{TableNumbers}. If the moments for D are scaled appropriately 
for a change of unit cell, the values given in 
Ref.~\protect\cite{a:ande88} are recovered. The moments for G and P are not
shown since they follow from those for D due to the 
Bonnet-transformation. The values for $\Delta$ are the same for G, D and P.}
\label{TableMoments}
\end{table}       

\end{document}